\documentclass[letterpaper,english,reprint, aps]{revtex4-1}
\usepackage[T1]{fontenc}
\usepackage[latin9]{inputenc}
\usepackage{color}
\usepackage{babel}
\usepackage{float}
\usepackage{amsmath}
\usepackage{amssymb}
\usepackage{graphicx}
\usepackage[unicode=true,pdfusetitle,
 bookmarks=true,bookmarksnumbered=false,bookmarksopen=false,
 breaklinks=false,pdfborder={0 0 1},backref=false,colorlinks=false]
 {hyperref}

\makeatletter

\pdfpageheight\paperheight
\pdfpagewidth\paperwidth

\usepackage[singlelinecheck=false,justification=RaggedRight,format=plain]{caption}

\@ifundefined{showcaptionsetup}{}{%
 \PassOptionsToPackage{caption=false}{subfig}}
\usepackage{subfig}
\makeatother

\begin{document}
\preprint{APS/123-QED}
\title{Probing Robust Majorana Signatures by Crossed Andreev Reflection with
a Quantum Dot}
\author{Guan-Hao Feng}
\author{Hong-Hao Zhang}
\email{zhh98@mail.sysu.edu.cn}

\affiliation{Guangdong Key Laboratory of Magnetoelectric Physics and Devices, School
of Physics, Sun Yat-sen University, Guangzhou 510275, China }
\begin{abstract}
We propose a three-terminal structure to probe robust signatures of
Majorana zero modes. This structure consists of a quantum dot coupled
to the normal metal, s-wave superconducting and Majorana Y-junction
leads. The zero-bias differential conductance at zero temperature
of the normal-metal lead peaks at $2e^{2}/h$, which will be deflected
after Majorana braiding. This quantized conductance can entirely arise
from the Majorana-induced crossed Andreev reflection, protected by
the energy gap of the superconducting lead. We find that the effect
of thermal broadening is significantly suppressed when the dot is
on resonance. In the case that the energy level of the quantum dot
is much larger than the superconducting gap, tunneling processes are
dominated by Majorana-induced crossed Andreev reflection. Particularly,
a novel kind of crossed Andreev reflection equivalent to the splitting
of charge quanta $3e$ occurs after Majorana braiding.
\end{abstract}
\maketitle

\section{INTRODUCTION}

Majorana zero modes (MZMs) are zero-energy quasiparticle excitations
originating from coherent superpositions of electrons and holes. Following
theoretical suggestions, MZMs are supported in 1D systems, such as
InAs or InSb wires with strong spin-orbit coupling and proximity-induced
superconductivity \citep{PhysRevLett.105.077001,PhysRevLett.105.177002},
and they show great potential in decoherence-free quantum computation
\citep{RN69,PhysRevLett.100.096407,PhysRevLett.105.227003,Kitaev_2001}.
Verifying the existence of MZMs and their non-Abelian braiding has
been attracting much attention in recent years \citep{PhysRevLett.98.237002,PhysRevLett.109.227006,PhysRevB.84.201308,PhysRevLett.102.216404,PhysRevLett.102.216403,PhysRevLett.124.036801,PhysRevLett.114.166406,PhysRevB.85.245121,PhysRevB.85.085415,PhysRevLett.122.097003}.

Due to the property that an MZM can act as both an electron lead and
a hole lead in tunneling processes, one of the most exciting theoretical
predictions is a quantized zero-bias conductance peak (ZBCP) of $\frac{2e^{2}}{h}$
at zero temperature \citep{PhysRevB.82.180516,PhysRevLett.103.237001,PhysRevLett.112.037001}.
However, it is quite difficult to observe this quantization from a
direct junction between a normal-metal lead and MZMs in a single-subband
wire because of thermal broadening, overlap of Majorana wave functions,
disorder, and localized Andreev bound states \citep{PhysRevLett.109.227005,PhysRevLett.109.227006,PhysRevLett.115.266804,PhysRevLett.109.267002,PhysRevLett.123.117001,PhysRevB.93.195161,PhysRevB.91.024514}.
Although the observation of ZBCP has been reported in many experiments
in recent years \citep{nadj2014observation,mourik2012signatures,RN70,Zhu189,PhysRevLett.116.257003},
the observation of MZMs has not been fully confirmed. Importantly,
very recently it has been recognized that one needs to be cautious
about the interpretation of non-quantized ZBCP as the signature of
MZMs in local tunneling experiments since such experiments only measure
one end of the one-dimensional setup \citep{frolov2021quantum}, while
the most important characteristics of MZMs are their nonlocal correlations.
To advance the pursuit of MZMs, new theoretical proposals and new
signatures which can reflect the nonlocal correlations of MZMs are
hence highly demanded. For example, shot noise and Fano factor in
Majorana setups can carry interesting information to identify MZMs
\citep{PhysRevLett.122.097003,PhysRevLett.98.237002,PhysRevLett.101.120403,PhysRevLett.116.166401,PhysRevB.91.081405,PhysRevB.83.153415,PhysRevB.85.085415,PhysRevB.97.155113,PhysRevB.99.165427,Smirnov_2017}.

\begin{figure}
\includegraphics[width=0.9\columnwidth]{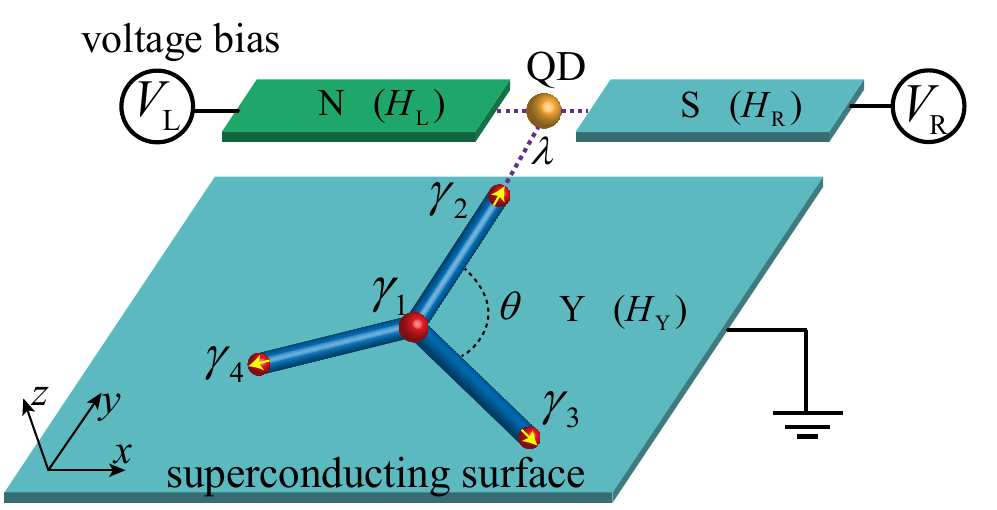}

\caption{\label{fig: Majorana-Y-junction}Setup of the T-shaped QD-(N, S, Y)
model with the normal-metal lead (N), the superconducting lead (S),
and the Majorana Y-junction lead (Y). Following Refs.\,\citep{van2012coulomb,PhysRevX.6.031019},
the Majorana braiding can be implemented on the Y junction by tuning
the couplings between the MZMs. }

\end{figure}

Here we propose a T-shaped hybrid structure to detect MZMs, as illustrated
in Fig.\,\ref{fig: Majorana-Y-junction}. The central quantum dot
(QD) acts as a transfer station of electrons and holes. Hence tunning
the energy level of the QD is equivalent to tunning the transmission
coefficients. The key to probe MZMs is the Majorana-induced crossed
Andreev reflection \citep{PhysRevLett.93.197003,PhysRevLett.103.237001,PhysRevLett.101.120403,PhysRevLett.122.257701}.
The ZBCP arising from the crossed Andreev reflection in this T-shaped
structure is strongly protected by the energy gap of the superconducting
lead because quasiparticle excitations are exponentially suppressed
$\sim\exp(-\Delta/T)$. Such kind of multiterminal structures with
a QD shows excellent maneuverability in the studies of spin-dependent
transport in strong Coulomb-correlated systems \citep{PhysRevLett.92.206801,PhysRevB.66.134507,PhysRevB.70.235341,PhysRevB.65.024516,PhysRevB.59.3831,PhysRevB.62.648,PhysRevB.78.155303}.

At zero temperature, we find that the ZBCP of the normal-metal lead
is quantized to $2e^{2}/h$ before braiding, which can be completely
induced by the crossed Andreev reflection. This quantized ZBCP is
found to be considerably robust against the temperature when the QD
is on resonance ($\epsilon_{\mathrm{d}}=0$). We show that the crossed
Andreev reflection dominates over the conventional Andreev reflection
when $\epsilon_{\mathrm{d}}\gg\Delta$. Importantly, we find that
the Majorana braiding shifts the ZBCPs and arouses a novel kind of
crossed Andreev reflection equivalent to the splitting of $3e$ charge
quanta, as shown in Fig.\,\ref{fig:Andreev reflection}. Because
of the high controllability of QD and the robustness of the predicted
signatures, our findings suggest a promising new way to identify MZMs.

\textcolor{black}{It is worth noting that while the Kondo correlations
are important in a strong coupling and low-temperature regime, the
Kondo resonances are usually either unstable or unquantized \citep{PhysRevB.82.245108,PhysRevX.4.031051,PhysRevB.92.235422}.
In sharp contrast, the Majorana-induced resonance in this paper is
always singly situated at zero bias and leads to quantized conductance.
In order to isolate and investigate observable consequences of the
Majorana-induced subgap resonances, we neglect the Kondo correlations
and focus on the Majorana-induced crossed Andreev reflection.}

This paper is organized as follows. In Sec.\,\ref{sec:MODEL-AND-FORMULATION},
we introduce the T-shaped hybrid model and explicitly write down their
Hamiltonians. In Sec.\,\ref{sec:CURRENT-AND-CONDUCTANCE}, we discuss
the electronic transport of the system and provide the corresponding
current and conductance formulas, including the analytical expressions
for the ZBCPs. In Sec.\,\ref{sec:SHOT-NOISE-AND}, we present the
formula for the shot noise and the Fano factor in terms of appropriate
Green's functions. The detailed derivations of the self-energy, the
local density of states, and the shot noise are given in Appendix
\ref{sec:CALCULATION OF THE SELF-ENERGY}, \ref{sec:CALCULATION-OF-LDOS}and
\ref{sec:CALCULATION OF THE SHOT NOISE}, respectively.

\section{MODEL AND FORMULATION\label{sec:MODEL-AND-FORMULATION}}

We introduce the three-terminal setup shown in Fig.\,\ref{fig: Majorana-Y-junction}.
The three leads are coupled with a central QD and the superconducting
lead ensures that the occurrence of crossed Andreev reflection, which
protects the ZBCP from quasiparticle excitations. The tunnel-coupled
structure can be described by an effective low-energy Hamiltonian:

\begin{equation}
H=H_{\mathrm{L}}+H_{\mathrm{R}}+H_{\mathrm{QD}}+H_{\mathrm{Y}}+H_{\mathrm{T}}.\label{eq:Effective Hamiltonian}
\end{equation}
The first term in Eq.\,\eqref{eq:Effective Hamiltonian} is the Hamiltonian
of the normal-metal lead (N) in Fig.\,\ref{fig: Majorana-Y-junction},
which is characterized by
\begin{equation}
H_{\mathrm{L}}=\sum_{k\sigma}\epsilon_{\mathrm{L},k\sigma}a_{\mathrm{L},k\sigma}^{\dagger}a_{\mathrm{L},k\sigma},\label{eq:left lead Hamiltonian}
\end{equation}
where $a_{\mathrm{L},k\sigma}^{\dagger}(a_{\mathrm{L},k\sigma})$
are creation (annihilation) operators with wave vector $k$ and spin
$\sigma=\uparrow,\downarrow$, and $\epsilon_{\mathrm{L},k\sigma}$
is the corresponding electron energy. The second term, the Hamiltonian
of the superconducting lead (S) in Fig.\,\ref{fig: Majorana-Y-junction},
is given by the BCS theory

\begin{equation}
H_{\mathrm{R}}=\sum_{k\sigma}\epsilon_{\mathrm{R},k\sigma}a_{\mathrm{R},k\sigma}^{\dagger}a_{\mathrm{R},k\sigma}+\sum_{k}(\Delta a_{\mathrm{R},k\uparrow}^{\dagger}a_{\mathrm{R},-k\downarrow}^{\dagger}+\mathrm{H.c.}).\label{eq:right lead Hamiltonian}
\end{equation}
The superconducting energy gap $\Delta$ is real here since a unitary
transformation  has been performed on this Hamiltonian \citep{PhysRevB.59.3831,PhysRevB.54.7366}.
In this work, we set the applied voltage of the superconducting lead
$V_{\mathrm{R}}=0$. For simplicity, we use the noninteracting Hamiltonian
of the QD

\begin{equation}
H_{\mathrm{QD}}=\sum_{\sigma}\epsilon_{\mathrm{d}}d_{\sigma}^{\dagger}d_{\sigma},
\end{equation}
where the QD level $\epsilon_{\mathrm{d}}=\epsilon_{0}-eV_{\mathrm{g}}/2$
is controlled by a gate voltage \textbf{$V_{\mathrm{g}}$} \citep{PhysRevB.78.155303,PhysRevB.82.245108,PhysRevB.84.201308}.
The Hamiltonian of the Majorana Y junction (Y) in Fig.\,\ref{fig: Majorana-Y-junction}
is given by

\begin{equation}
H_{\mathrm{Y}}=i\sum_{k=2}^{4}t_{1k}\gamma_{1}\gamma_{k},\label{eq:Hamiltonian of the Y junction}
\end{equation}
where the Coulomb coupling constants are $t_{12}=t_{13}=t_{\mathrm{\min}}$
and $t_{14}=t_{\mathrm{\max}}$ with $t_{\mathrm{\min}}\ll t_{\max}$
\citep{van2012coulomb}. Using two fermionic operators $c_{1}=(\gamma_{1}-i\gamma_{4})/2$
and $c_{2}=(\gamma_{2}-i\gamma_{3})/2$, the Hamiltonian  $H_{\mathrm{Y}}$
can be represented in the four-dimensional Nambu-spinor space spanned
by $c_{\mathrm{Y}}^{\dagger}=(c_{1}^{\dagger},c_{1},c_{2}^{\dagger},c_{2})$.

\textcolor{black}{The tunneling Hamiltonian consists of}

\textcolor{black}{
\begin{equation}
H_{\mathrm{T}}=H_{\mathrm{T,L}}+H_{\mathrm{T,R}}+H_{\mathrm{T,Y}},
\end{equation}
where}

\textcolor{black}{
\begin{equation}
H_{\mathrm{T,L(R)}}=\sum_{k\sigma}v_{\mathrm{L}(\mathrm{R}),k}d_{\sigma}^{\dagger}a_{\mathrm{L(R)},k\sigma}+\mathrm{H.c.},
\end{equation}
}with $v_{\mathrm{L},k}$ and $v_{\mathrm{R},k}$ denoting the complex
tunneling amplitudes of the normal-metal and superconducting leads,
respectively. The coupling between the QD and the Majorana lead is
spin-conserving, \emph{i.e.}, the MZM is always tunnel-coupled to
electrons in the QD with the same spin orientation \citep{PhysRevB.94.014502}.
Since we have set the spin orientation of the Rashba spin-orbit coupling
along the z-direction in Fig.\,\ref{fig: Majorana-Y-junction}, the
spin of each MZM (except $\gamma_{1}$) is parallel to the axial direction
of the corresponding nanowire \citep{PhysRevLett.105.177002,Xu_2017}.
Defining that the spin-$\uparrow$ direction is along the y-direction,
the coupling between the QD and the Majorana lead is given by

\begin{equation}
H_{\mathrm{T,Y}}=\lambda d_{\uparrow}^{\dagger}\gamma_{2}+\mathrm{H.c.},\label{eq:Y junction tunneling}
\end{equation}
where $\lambda$ is the coupling amplitude. For simplicity, we assume
$\lambda$ is real.

\section{CURRENT AND CONDUCTANCE\label{sec:CURRENT-AND-CONDUCTANCE}}

The ZBCP arising from the crossed Andreev reflection in this T-shaped
structure is a remarkable signature of MZMs. In this section, we first
calculate the time-average current by using the nonequilibrium Green's
function method \citep{PhysRevLett.68.2512,PhysRevB.85.085415,haug2008quantum,PhysRevB.54.7366,PhysRevB.68.115319,keldysh1965diagram},
and then derive the analytic expression of the ZBCP of each lead.

The time-average current of the normal-metal lead is given by

\begin{align}
I_{\mathrm{L}} & =-e\bigl\langle\dot{N}_{\mathrm{L}}(t)\bigr\rangle\nonumber \\
 & =\frac{e}{h}\int\mathrm{d}\omega\thinspace\mathrm{Re}\thinspace\mathrm{Tr}\thinspace\{[G_{\mathrm{QD}}^{R}(\omega)\Sigma_{\mathrm{L}}^{<}(\omega)+G_{\mathrm{QD}}^{<}(\omega)\Sigma_{\mathrm{L}}^{A}(\omega)]\widetilde{\sigma}_{z}\},\label{eq:Current in GF form}
\end{align}
where $N_{L}(t)=\sum_{k\sigma}a_{\mathrm{L},k\sigma}^{\dagger}(t)a_{\mathrm{L},k\sigma}(t)$
is the total number operator of the electrons in the normal-metal
lead . The $4\times4$ Green's functions $G_{\mathrm{QD}}^{<}(t,t')\equiv-i\left\langle d(t')d^{\dagger}(t)\right\rangle $
and $G_{\mathrm{QD}}^{R}(t,t')\equiv-i\theta(t-t')\left\langle \left\{ d(t),d^{\dagger}(t')\right\} \right\rangle $
is defined with the Nambu spinors $d^{\dagger}=(d_{\uparrow}^{\dagger},d_{\downarrow},d_{\downarrow}^{\dagger},d_{\uparrow})$.
The retarded self-energy $\Sigma_{\mathrm{L}}^{R}(\omega)=[\Sigma_{L}^{A}(\omega)]^{\dagger}=\sum_{k}\mathcal{H}_{\mathrm{T,L}}^{\dagger}g_{\mathrm{L}}^{R}(\omega)\mathcal{H}_{\mathrm{T,L}}$
is defined with the Nambu spinors $a_{\mathrm{L(R)}}^{\dagger}=(a_{\mathrm{L(R)},k\uparrow}^{\dagger},a_{\mathrm{L(R)},-k\downarrow},a_{\mathrm{L(R)},-k\downarrow}^{\dagger},a_{\mathrm{L(R)},k\uparrow})$.
Here $g_{\mathrm{L}}^{R}(\omega)=(\omega-\mathcal{H}_{\mathrm{L}}+i0^{+})^{-1}$
is the corresponding unperturbed Keldysh contour Green's functions
of the normal-metal lead. The lesser self-energy is $\Sigma_{\mathrm{L}}^{<}(\omega)=F_{\mathrm{L}}(\Sigma_{\mathrm{L}}^{A}(\omega)-\Sigma_{\mathrm{L}}^{R}(\omega))$,
where $F_{\mathrm{L}}=\mathrm{diag}(f_{\mathrm{L}},\bar{f}_{\mathrm{L}},f_{\mathrm{L}}\bar{f}_{\mathrm{L}})$
is the Fermi distribution function matrix with $f_{\mathrm{L}}=f(\omega-eV_{\mathrm{L}})$
and $\bar{f}_{\mathrm{L}}=f(\omega+eV_{\mathrm{L}})$. The matrix
$\widetilde{\sigma}_{z}=\mathrm{diag}(1,-1,1,-1)$ describes the different
charge of electrons and holes.

The time-average current Eq.\,\eqref{eq:Current in GF form} is calculated
in terms of $\Sigma_{\mathrm{L}}^{R,A,<}(\omega)$ and $G_{\mathrm{QD}}^{R,A,<}(\omega)$.
This expression can be generalized to $I_{\mathrm{\eta}}$ by replacing
the self-energies $\Sigma_{\mathrm{L}}^{R,A,<}(\omega)$ with $\Sigma_{\mathrm{\eta}}^{R,A,<}(\omega)$
for $\eta=\mathrm{L},\mathrm{R}$ and $\mathrm{Y}$ representing the
normal-metal lead, the superconducting lead and the Majorana lead,
respectively.\textcolor{black}{{} In the basis $(d^{\dagger},a_{\mathrm{L}}^{\dagger},a_{\mathrm{R}}^{\dagger},c_{\mathrm{Y}}^{\dagger})$,
the Hamiltonian Eq.\,\eqref{eq:Effective Hamiltonian} can be written
in a block form as}

\textcolor{black}{
\begin{equation}
\mathcal{H}=\left(\begin{array}{cccc}
\mathcal{H}_{\mathrm{QD}} & \mathcal{H}_{\mathrm{T,L}} & \mathcal{H}_{\mathrm{T,R}} & \mathcal{H}_{\mathrm{T,Y}}\\
\mathcal{H}_{\mathrm{T,L}}^{\dagger} & \mathcal{H}_{\mathrm{L}} & 0 & 0\\
\mathcal{H}_{\mathrm{T,R}}^{\dagger} & 0 & \mathcal{H}_{\mathrm{R}} & 0\\
\mathcal{H}_{\mathrm{T,Y}}^{\dagger} & 0 & 0 & \mathcal{H}_{\mathrm{Y}}
\end{array}\right),
\end{equation}
}where the sub-matrices are given by
\begin{subequations}
\begin{equation}
\mathcal{H}_{\mathrm{L}}=\left(\begin{array}{cccc}
\epsilon_{\mathrm{L,k\uparrow}} & 0 & 0 & 0\\
0 & -\epsilon_{\mathrm{L,k\downarrow}} & 0 & 0\\
0 & 0 & \epsilon_{\mathrm{L,k\downarrow}} & 0\\
0 & 0 & 0 & -\epsilon_{\mathrm{L},k\uparrow}
\end{array}\right),
\end{equation}

\begin{equation}
\mathcal{H}_{\mathrm{R}}=\left(\begin{array}{cccc}
\epsilon_{\mathrm{R,k\uparrow}} & \Delta & 0 & 0\\
\Delta & -\epsilon_{\mathrm{R,k\downarrow}} & 0 & 0\\
0 & 0 & \epsilon_{\mathrm{R,k\downarrow}} & -\Delta\\
0 & 0 & -\Delta & -\epsilon_{\mathrm{R},k\uparrow}
\end{array}\right),
\end{equation}

\begin{equation}
\mathcal{H}_{\mathrm{QD}}=\left(\begin{array}{cccc}
\epsilon_{\mathrm{d}} & 0 & 0 & 0\\
0 & -\epsilon_{\mathrm{d}} & 0 & 0\\
0 & 0 & \epsilon_{\mathrm{d}} & 0\\
0 & 0 & 0 & -\epsilon_{\mathrm{d}}
\end{array}\right),
\end{equation}

\begin{equation}
\mathcal{H}_{\mathrm{Y}}=\left(\begin{array}{cccc}
-2t_{14} & 0 & it_{\mathrm{12}}-t_{13} & it_{\mathrm{12}}+t_{13}\\
0 & 2t_{\mathrm{14}} & it_{\mathrm{12}}-t_{13} & it_{\mathrm{12}}+t_{13}\\
-it_{\mathrm{12}}-t_{13} & -it_{\mathrm{12}}-t_{13} & 0 & 0\\
-it_{\mathrm{12}}+t_{13} & -it_{\mathrm{12}}+t_{13} & 0 & 0
\end{array}\right),
\end{equation}

\begin{equation}
\mathcal{H}_{\mathrm{T,L(R)}}=\left(\begin{array}{cccc}
v_{\mathrm{L(R)},k} & 0 & 0 & 0\\
0 & -v_{\mathrm{L(R)},k}^{*} & 0 & 0\\
0 & 0 & v_{\mathrm{L(R)},k} & 0\\
0 & 0 & 0 & -v_{\mathrm{L(R)},k}^{*}
\end{array}\right),
\end{equation}

\begin{equation}
\mathcal{H}_{\mathrm{T,Y}}=\left(\begin{array}{cccc}
0 & 0 & \lambda & \lambda\\
0 & 0 & 0 & 0\\
0 & 0 & 0 & 0\\
0 & 0 & -\lambda & -\lambda
\end{array}\right).
\end{equation}
\end{subequations}

Assuming that the electron energies in Eqs.\,\ref{eq:left lead Hamiltonian}
and\,\ref{eq:right lead Hamiltonian} are independent of spins with
$\epsilon_{\mathrm{L(R)},k\uparrow}=\epsilon_{\mathrm{L(R)},k\downarrow}=\epsilon_{\mathrm{L(R)},k}$
, the retarded self-energies from the couplings between the three
leads and the QD are given by
\begin{align}
\Sigma_{\mathrm{L}}^{R}(\omega) & =-\frac{i}{2}\Gamma_{\mathrm{L}}\left(\begin{array}{cccc}
1 & 0 & 0 & 0\\
0 & 1 & 0 & 0\\
0 & 0 & 1 & 0\\
0 & 0 & 0 & 1
\end{array}\right),\\
\Sigma_{\mathrm{R}}^{R}(\omega) & =-\frac{i}{2}\Gamma_{\mathrm{R}}\beta(\omega)\left(\begin{array}{cccc}
1 & -\frac{\Delta}{\omega} & 0 & 0\\
-\frac{\Delta}{\omega} & 1 & 0 & 0\\
0 & 0 & 1 & \frac{\Delta}{\omega}\\
0 & 0 & \frac{\Delta}{\omega} & 1
\end{array}\right),\\
\Sigma_{\mathrm{Y}}^{R}(\omega) & =\kappa\Lambda=\kappa\left(\begin{array}{cccc}
1 & 0 & 0 & -1\\
0 & 0 & 0 & 0\\
0 & 0 & 0 & 0\\
-1 & 0 & 0 & 1
\end{array}\right),\label{eq:self-energy of the Majorana lead}
\end{align}
where $\Gamma_{\mathrm{L(R)}}=2\pi|\nu_{\mathrm{L(R)},k}|^{2}\rho_{\mathrm{L(R)}}$
is the line-width function with $\rho_{\mathrm{L(R)}}$ being the
density of normal states. In the wide-band limit, $\Gamma_{\mathrm{L(R)}}$
is a constant independent of the frequency $\omega$. Here $\beta(\omega)=\frac{|\omega|\Theta(|\omega|-\Delta)}{\sqrt{\omega^{2}-\Delta^{2}}}+\frac{\omega\Theta(\Delta-|\omega|)}{i\sqrt{\Delta^{2}-\omega^{2}}}$
is the dimensionless BCS density of states. Neglecting the contribution
of $t_{\min}$, we have $\kappa\thickapprox\frac{-8\lambda^{2}t_{\max}^{2}+2\lambda^{2}(\omega^{+})^{2}}{-4t_{\mathrm{\max}}^{2}\omega^{+}+(\omega^{+})^{3}}$
with $\omega^{+}=\omega+i0^{+}$.

The lesser Green's function of the QD can be obtained via the Keldysh
equation $G_{\mathrm{QD}}^{<}(\omega)=G_{\mathrm{QD}}^{R}(\omega)\Sigma_{\mathrm{TOT}}^{<}(\omega)G_{\mathrm{QD}}^{A}(\omega)$,
with $\Sigma_{\mathrm{TOT}}^{<}=\sum_{\eta}\Sigma_{\mathrm{\eta}}^{<}$
. In the calculation of the time-average current, we can take $\Sigma_{\mathrm{Y}}^{<}=0$
(see Appendix \ref{sec:CALCULATION OF THE SELF-ENERGY} for details).
In the case of $|eV_{\mathrm{L}}|<|\Delta|$, analytical calculations
for the time-average current yield $I_{\eta}=\frac{e}{2h}\int\mathrm{d}\omega(f_{\mathrm{L}}-\bar{f_{\mathrm{L}}})T_{\eta}(\omega)$,
which means that only Andreev reflection processes contribute to the
electronic transport of the system. Since the system is in a stationary
regime, the total current is conserved, \emph{i.e.}, $\sum_{\eta}I_{\eta}=0$.
\begin{figure*}
\noindent
\subfloat[\label{fig:Conductance of the left lead 1}]{\includegraphics[width=0.65\columnwidth]{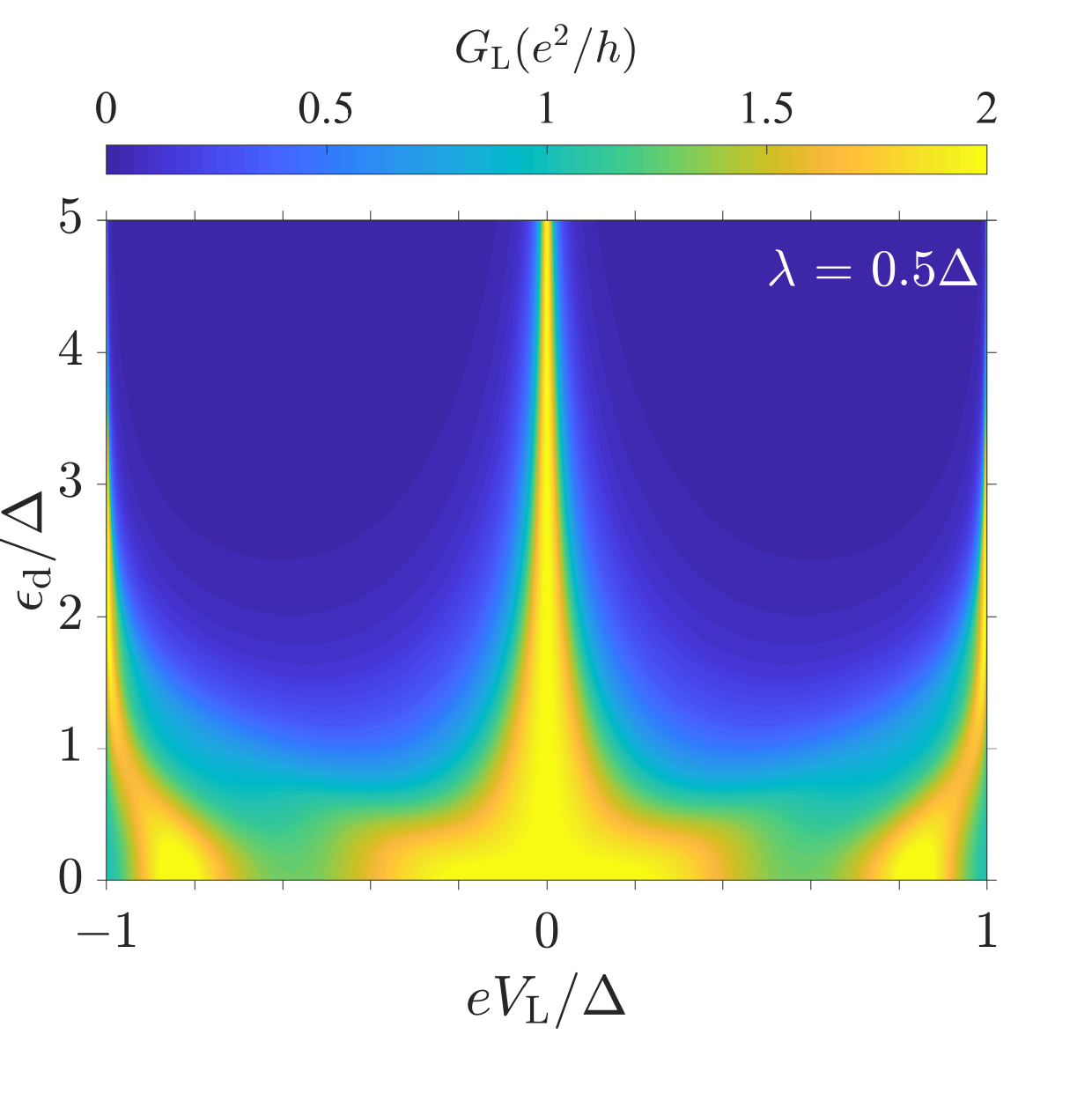}}
\subfloat[\label{fig:Conductance of the left lead 2}]{\includegraphics[width=0.65\columnwidth]{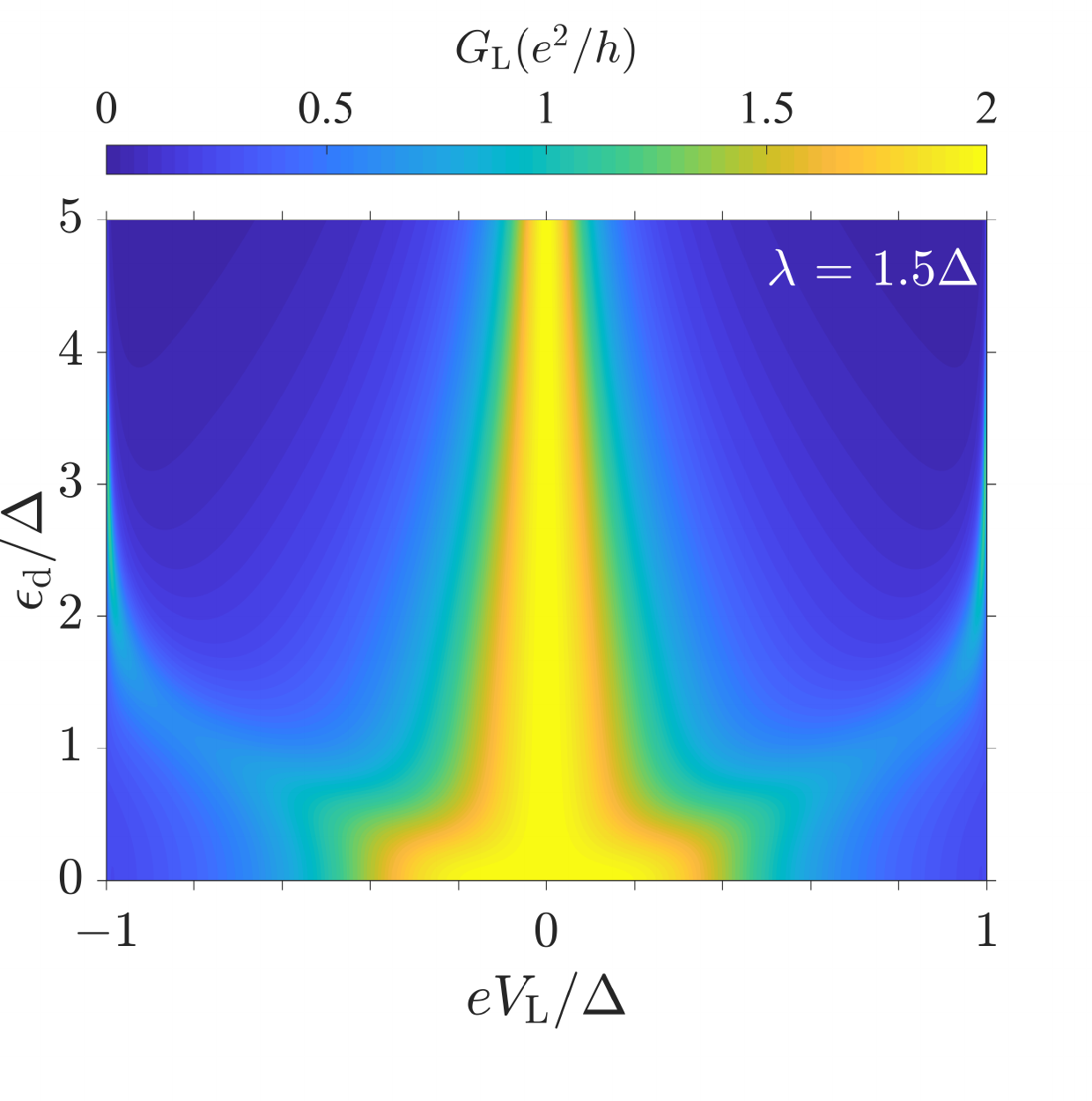}}
\subfloat[\label{fig:Conductance of the left lead 3}]{\includegraphics[width=0.65\columnwidth]{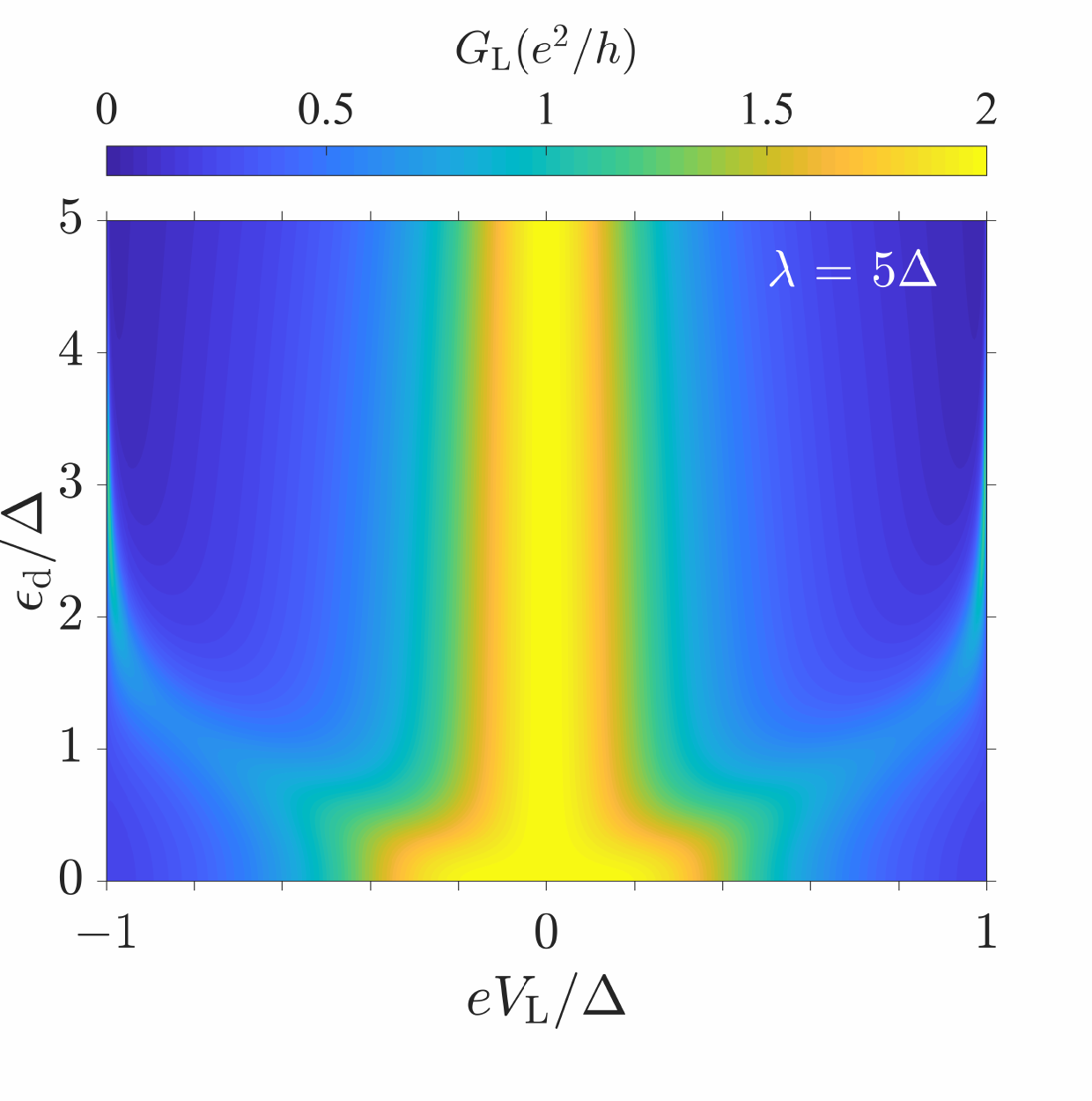}}

\noindent
\subfloat[\label{fig:LDOS 1}]{\includegraphics[viewport=55bp 0bp 640bp 600bp,clip,width=0.63\columnwidth,height=0.6\columnwidth]{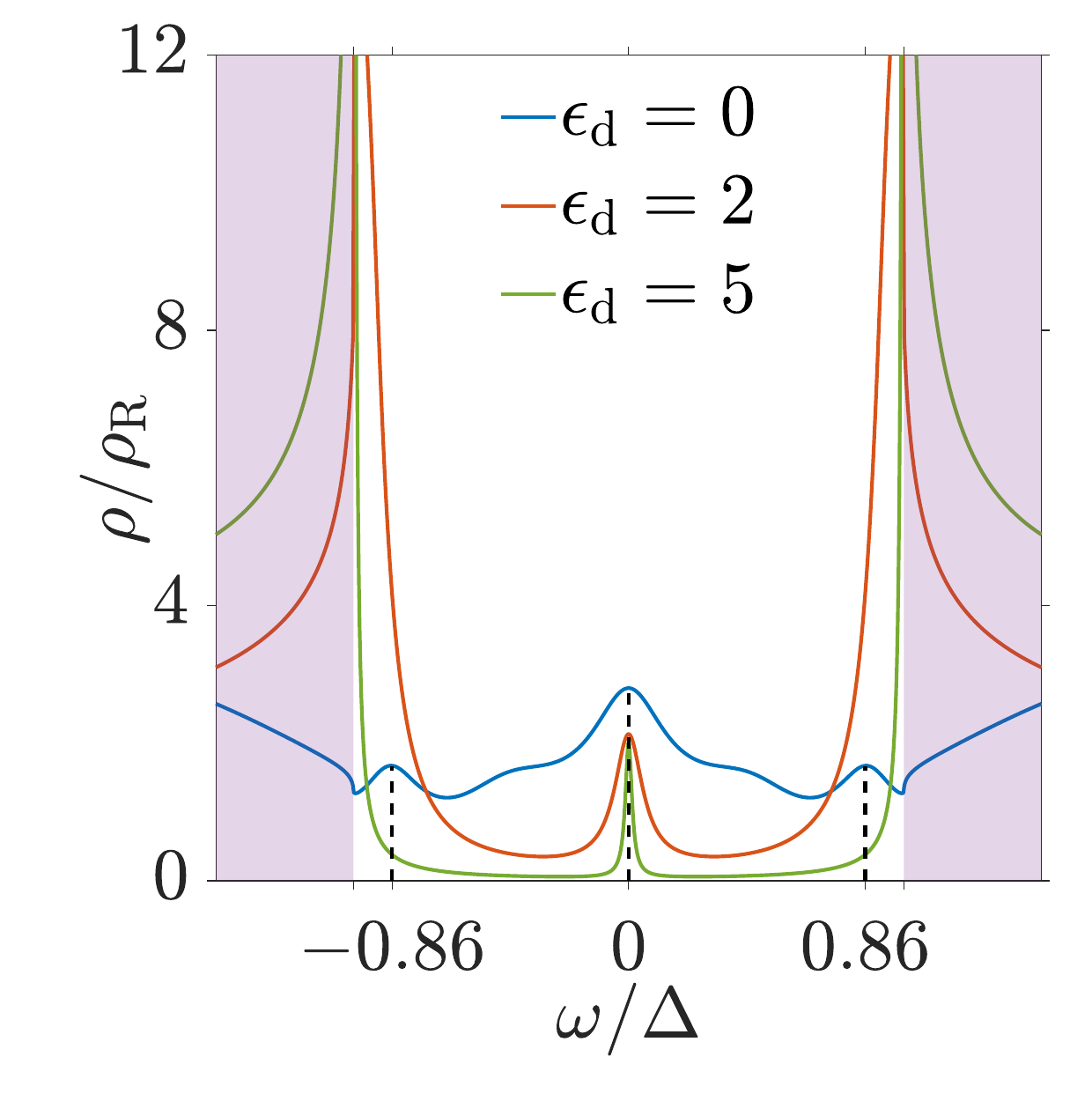}}
\subfloat[\label{fig:LDOS 2}]{\includegraphics[viewport=55bp 0bp 610bp 600bp,clip,width=0.63\columnwidth,height=0.6\columnwidth]{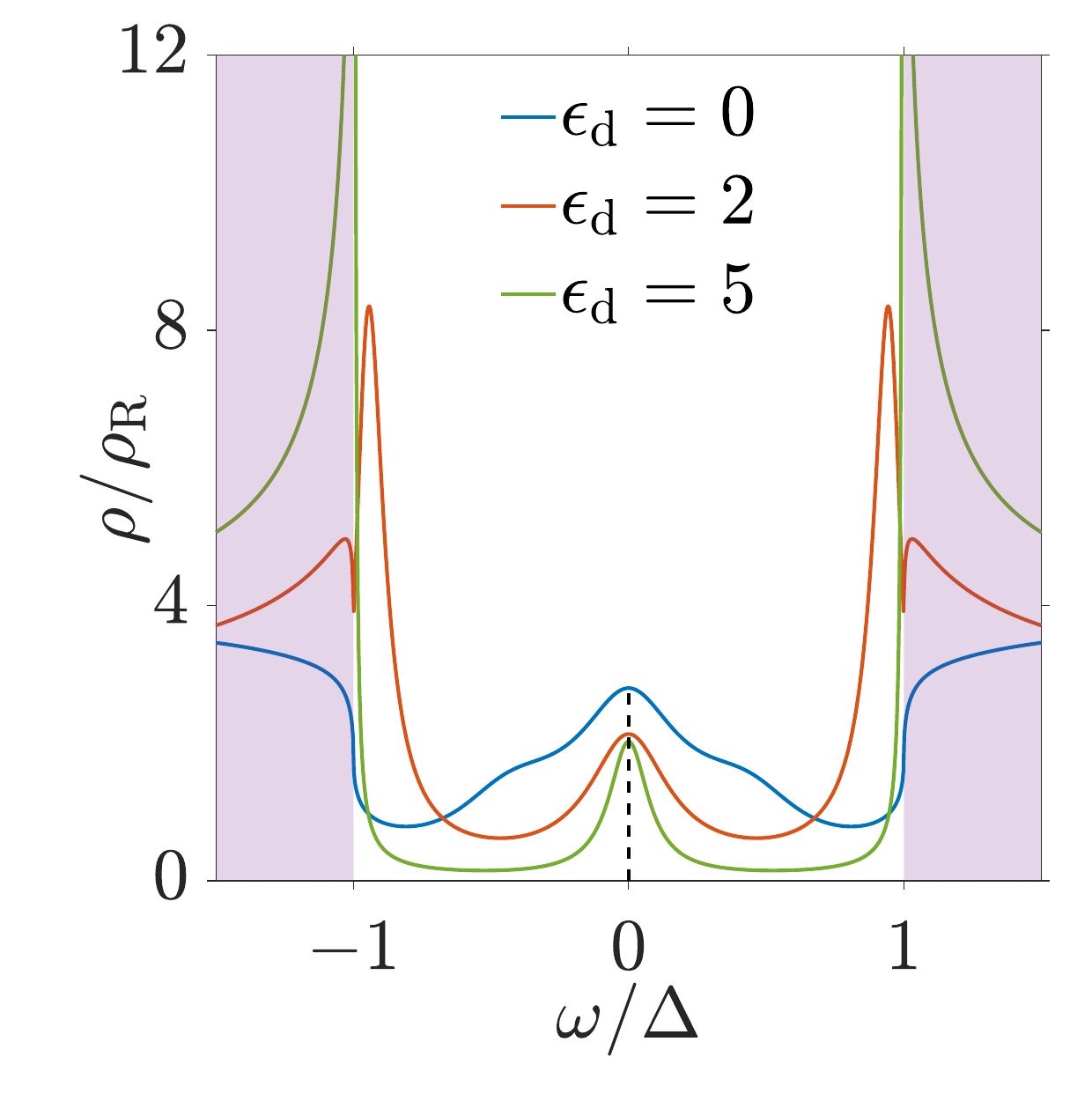}}
\subfloat[\label{fig:LDOS 3}]{\includegraphics[viewport=30bp 0bp 600bp 600bp,clip,width=0.63\columnwidth,height=0.6\columnwidth]{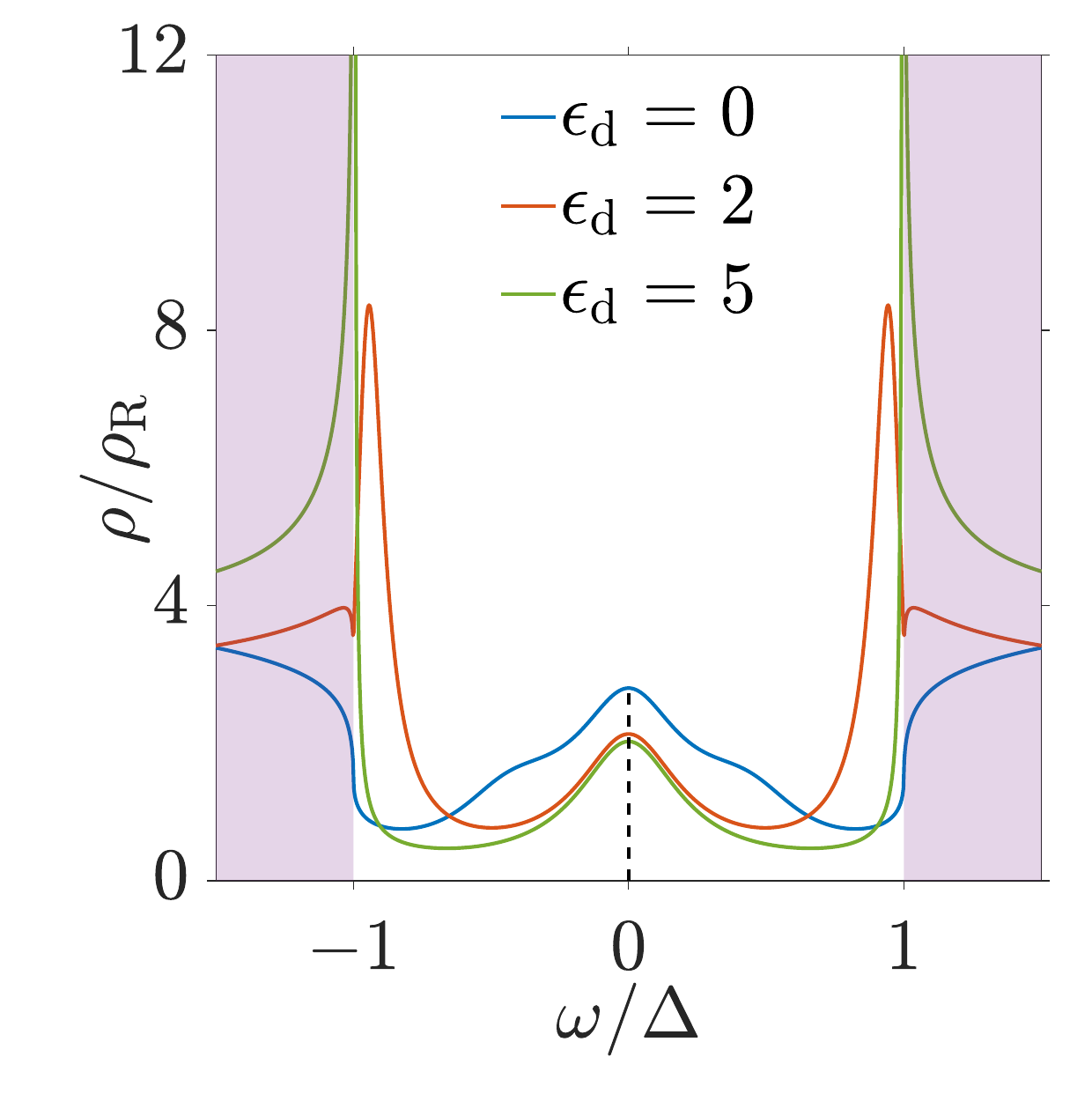}}
\caption{\label{fig:The differential conductances} (a)-(c) Differential conductance
spectra of the normal-metal lead as functions of $eV_{\mathrm{L}}/\Delta$
and $\epsilon_{\mathrm{d}}/\Delta$ at zero temperature. The parameters
are $\Gamma_{\mathrm{L}}=\Gamma_{\mathrm{R}}=0.8\Delta$, $t_{\min}=0.001\Delta$,
and $t_{\max}=\Delta$. \textcolor{black}{(e)-(f) The LDOS of the
superconducting lead with the same parameters in (a)-(c). The LDOS
is defined by the sum of the diagonal spectral function of the superconducting
lead, }\textcolor{black}{\emph{i.e.}}\textcolor{black}{, $\rho(\omega)=-\mathrm{Im}\left[\mathrm{Tr}\thinspace G_{\mathrm{R}}^{R}\right]/\pi$.}}
\end{figure*}

 The differential conductances of the leads $\eta$ at zero temperature
are obtained by $G_{\eta}=dI_{\eta}/dV_{\mathrm{L}}$. Especially,
the ZBCPs at zero temperature are

\begin{align}
\lim_{eV_{\mathrm{L}}\to0}G_{\mathrm{L}}(eV_{\mathrm{L}}) & =\begin{cases}
\frac{2e^{2}}{h}, & \lambda\neq0,\\
\frac{e^{2}}{h}\frac{16\Gamma_{\mathrm{L}}^{2}\Gamma_{\mathrm{R}}^{2}}{(\Gamma_{\mathrm{L}}^{2}+\Gamma_{\mathrm{R}}^{2}+4\epsilon_{\mathrm{d}}^{2})^{2}}, & \lambda=0,
\end{cases}\label{eq:zero-bias left lead conductance}\\
\lim_{eV_{\mathrm{L}}\to0}G_{\mathrm{R}}(eV_{\mathrm{L}}) & =\begin{cases}
-\frac{e^{2}}{h}\frac{4\Gamma_{\mathrm{R}}^{2}}{\Gamma_{\mathrm{L}}^{2}+\Gamma_{\mathrm{R}}^{2}+4\epsilon_{\mathrm{d}}^{2}}, & \lambda\neq0,\\
-\frac{e^{2}}{h}\frac{16\Gamma_{\mathrm{L}}^{2}\Gamma_{\mathrm{R}}^{2}}{(\Gamma_{\mathrm{L}}^{2}+\Gamma_{\mathrm{R}}^{2}+4\epsilon_{\mathrm{d}}^{2})^{2}}, & \lambda=0,
\end{cases}\label{eq:zero-bias right lead conductance}\\
\lim_{eV_{\mathrm{L}}\to0}G_{\mathrm{Y}}(eV_{\mathrm{L}}). & =\begin{cases}
-\frac{e^{2}}{h}\frac{2(\Gamma_{\mathrm{L}}^{2}-\Gamma_{\mathrm{R}}^{2}+4\epsilon_{\mathrm{d}}^{2})}{\Gamma_{\mathrm{L}}^{2}+\Gamma_{\mathrm{R}}^{2}+4\epsilon_{\mathrm{d}}^{2}}, & \lambda\neq0,\\
0, & \lambda=0.
\end{cases}\label{eq:zero-bias Y lead conductance}
\end{align}

When $\lambda=0$, the Majorana Y junction is disconnected with the
QD, and the remaining part is reduced to an N-QD-S structure. The
maximal ZBCP in Eq.\,\eqref{eq:zero-bias left lead conductance}
is equal to $4e^{2}/h$ when the QD is symmetrically coupled $(\Gamma_{\mathrm{L}}=\Gamma_{\mathrm{R}})$
and on resonance $(\epsilon_{\mathrm{d}}=0)$, in accord with the
previous results of Ref.\,\citep{PhysRevB.70.235341}. When $\lambda\neq0$,
the ZBCP of the normal-metal lead in this three-terminal structure
equals a quantized value $2e^{2}/h$, which is consistent with the
famous conductance peak for the N-TS tunneling. As depicted in Figs.\,\ref{fig:Conductance of the left lead 1}-\ref{fig:Conductance of the left lead 3},
the ZBCP is obviously broad for $\epsilon_{\mathrm{d}}=0$ and becomes
sharp for large $\epsilon_{\mathrm{d}}$. The QD acts as a transfer
station of electrons and holes, which means that the energy level
of the QD is the tunnel barrier of the system. Hence the broadening
of the ZBCP arises from the junction transparency effect and the height
of the ZBCP is not affected. This quantized ZBCP is caused by the
perfect Majorana-induced Andreev reflection. In the next section,
we will show that the local Andreev reflection can be completely suppressed
by increasing $\epsilon_{\mathrm{d}}$, and only the crossed Andreev
reflection remains. We emphasize again that this ZBCP of $2e^{2}/h$
can completely arise from the crossed Andreev reflection, which is
strongly protected by the superconducting gap $\Delta$ \citep{PhysRevLett.115.266804}.
Moreover, the results of Eqs.\,\ref{eq:zero-bias right lead conductance}
and\,\ref{eq:zero-bias Y lead conductance} show that the ZBCPs of
both the superconducting and the Majorana leads are insensitive to
the nonzero coupling amplitude $\lambda$, but only dependent on $\Gamma_{\mathrm{L}}$,
$\Gamma_{R}$ and $\epsilon_{\mathrm{d}}$.

\textcolor{black}{The conductance peaks of the normal-metal lead are
closely related to the local density of states (LDOS) of the superconducting
lead in this T-shaped structure (the details of the analytical derivation
of the LDOS are provided in Appendix\,\ref{sec:CALCULATION-OF-LDOS}).
As shown in Figs.\,\ref{fig:LDOS 1}-\ref{fig:LDOS 3}, there are
three subgap resonances in the superconducting lead two are the spin-induced
resonances situating near the gap edge \citep{PhysRevB.82.245108},
one is the Majorana-induced resonance situating at $\omega=0$. The
conductance peaks are all situated at the subgap resonance energy.
As $\epsilon_{\mathrm{d}}$ increases, all the resonances become sharper,
and the two near the gap edge merge with the dips eventually, whereas
the one at $\omega=0$ remains. For large $\epsilon_{\mathrm{d}}$,
the Majorana-induced resonance situating at $\omega=0$ sharpens to
form a localized bound state, }\textcolor{black}{\emph{i.e.}}\textcolor{black}{,
a Yu-Shiba-Rusinov state (YSR state). The occurrence of the Majorana-induced
YSR state will lead to the domination of crossed Andreev reflection,
which will be discussed in the next section. The Majorana-induced
resonance in the superconducting lead is not quantized but parameters
dependent, consistent with the result in Eq.\,\ref{eq:zero-bias right lead conductance}.
The Majorana-induced resonance leads to a quantized ZBCP of the normal-metal
lead since the electrons and holes are transported through perfect
Andreev reflection \citep{PhysRevB.103.214502}. The spin-induced
conductance peaks are unstable and unquantized, }\textcolor{black}{\emph{i.e}}\textcolor{black}{.,
they are not robust as functions of parameters. When $\lambda$ increases,
the Majorana induced resonance is enhanced and the spin-induced conductance
peaks become inconspicuous in Figs.\,\ref{fig:Conductance of the left lead 2}
and \ref{fig:Conductance of the left lead 3} due to the competition
between the spin-induced and the Majorana-induced resonances in the
tunneling processes.}

\noindent
\begin{figure}
\includegraphics[width=1\columnwidth]{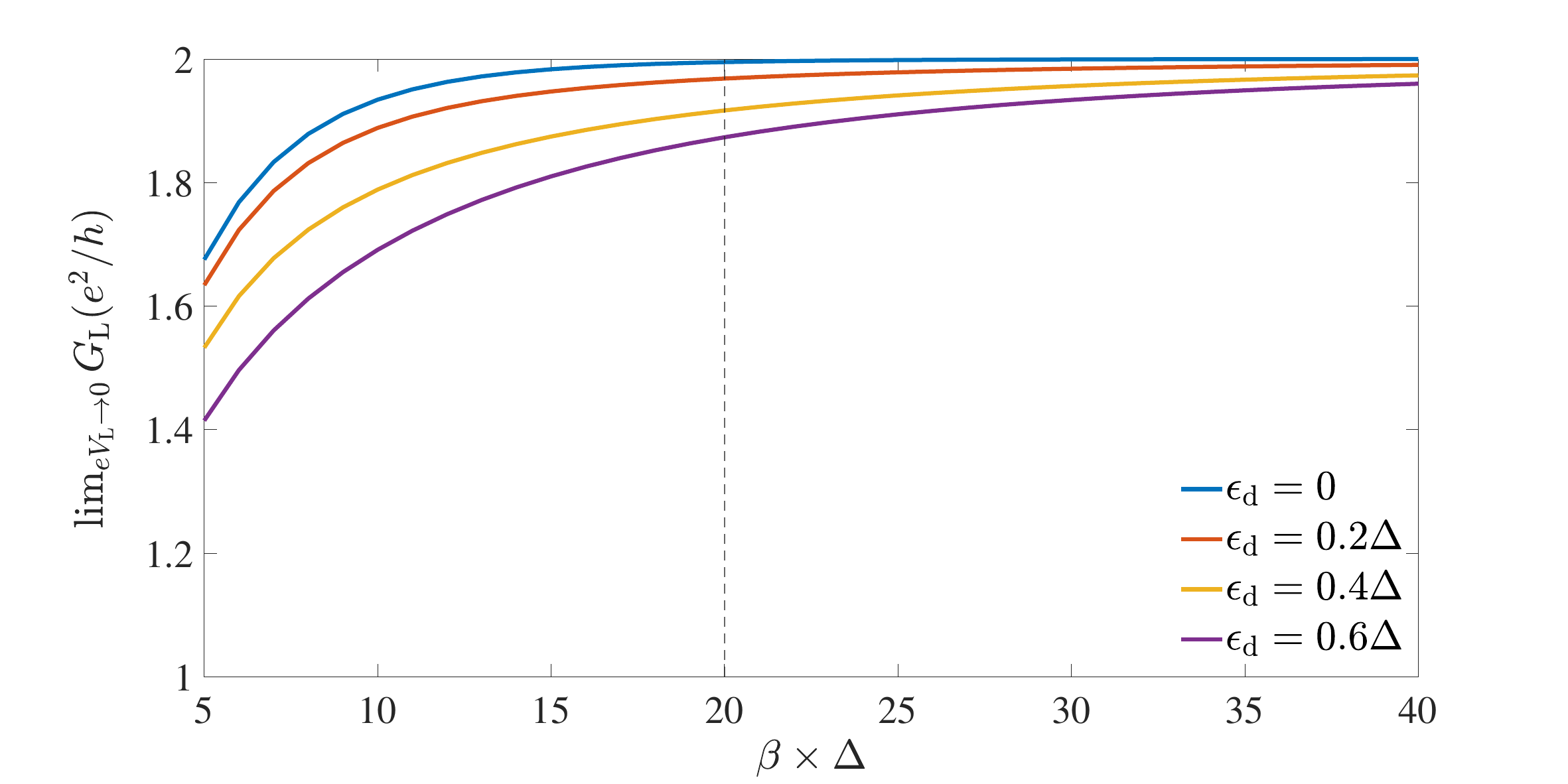}\caption{\label{fig: Plot of peak vs temperature}ZBCP of the normal-metal
lead as a function of finite temperature. The parameters are $\Gamma_{\mathrm{L}}=\Gamma_{\mathrm{R}}=0.8\Delta$,
$\lambda=\Delta$, $t_{\min}=10^{-3}\Delta$, and $t_{\max}=\Delta$.}
\end{figure}

The discussion can easily extend to finite temperature regimes. As
shown in Fig.\,\ref{fig: Plot of peak vs temperature}, the ZBCP
of the normal-metal lead is no longer quantized to $2e^{2}/h$, since
the Fermi distribution is smoothly dependent on the temperature $T$,
which is called the thermal broadening. Nevertheless, we find that
the effect of the thermal broadening is significantly suppressed by
large junction transparency $(\epsilon_{\mathrm{d}}=0$\emph{). }In
Fig.\,\ref{fig: Plot of peak vs temperature}, the ZBCP is pretty
close to $2e^{2}/h$ when $k_{B}T<\Delta/20$. Such a temperature
condition can be met in the experiment, e.g., see Ref.\,\citep{mourik2012signatures},
in which the induced superconducting gap of the InSb nanowires is
$\Delta\approx250\mathrm{\mu eV}$ and the minimized temperature is
$k_{B}T\approx4.3\mathrm{\mu eV}$.

\section{SHOT NOISE AND FANO FACTOR\label{sec:SHOT-NOISE-AND}}

In addition to the time-average current, the shot noise can reveal
the fluctuation of the current and provide useful information about
MZMs \citep{PhysRevLett.98.237002,PhysRevLett.101.120403,PhysRevB.83.153415,PhysRevB.79.161408}.
The shot noise, defined as the correlation function of the current
fluctuations between leads $\eta$ and $\eta'$, takes the form $S_{\eta\eta'}(t,t')=\left\langle \{\delta I_{\eta}(t),\delta I_{\eta'}(t')\}\right\rangle $,
where $\delta I_{\eta}(t)=\hat{I}_{\eta}(t)-I_{\eta}$, and $\hat{I}_{\eta}(t)=-e\dot{N}_{\eta}(t)$.
The time-average current $I_{\eta}$ has been obtained by Eq.\,\eqref{eq:Current in GF form}.
With the use of the Wick's theorem and the S-matrix expansion \citep{PhysRevB.78.155303,PhysRevB.85.085415},
we can reduce the expression of the shot noise in terms of Green's
functions. After Fourier transform, we obtain the expression of shot
noise in the frequency space $S_{\eta\eta'}(\omega')$. The calculation
of the shot noise is shown explicitly in Appendix \ref{sec:CALCULATION OF THE SHOT NOISE}.

Following Ref.\, \citep{BLANTER20001}, in multi-terminal systems,
the shot noise $S_{\eta\eta'}(\omega')$ with $\eta=\eta'$ must be
positive; conversely, that with $\eta\neq\eta'$ must be negative.
This property can be verified by the numerical calculation of $S_{\eta\eta}(0)$
in the following. The zero-frequency Fano factor, defined by the ratio
$F_{\eta}=S_{\eta\eta}(0)/2eI_{\eta}$, can gain insight into the
nature of charge quanta transferred to lead $\eta$ \citep{PhysRevB.83.153415,PhysRevLett.86.4104,PhysRevLett.82.4086}.\textcolor{red}{\emph{
}}\textcolor{black}{Beyond the linear regime in this paper, the Majorana-induced
nonlinear effective charge has been studied in detail \citep{Smirnov_2017},
which gives rise to fractional effective charge quanta.}

\begin{figure}[h]
\includegraphics[width=0.95\columnwidth]{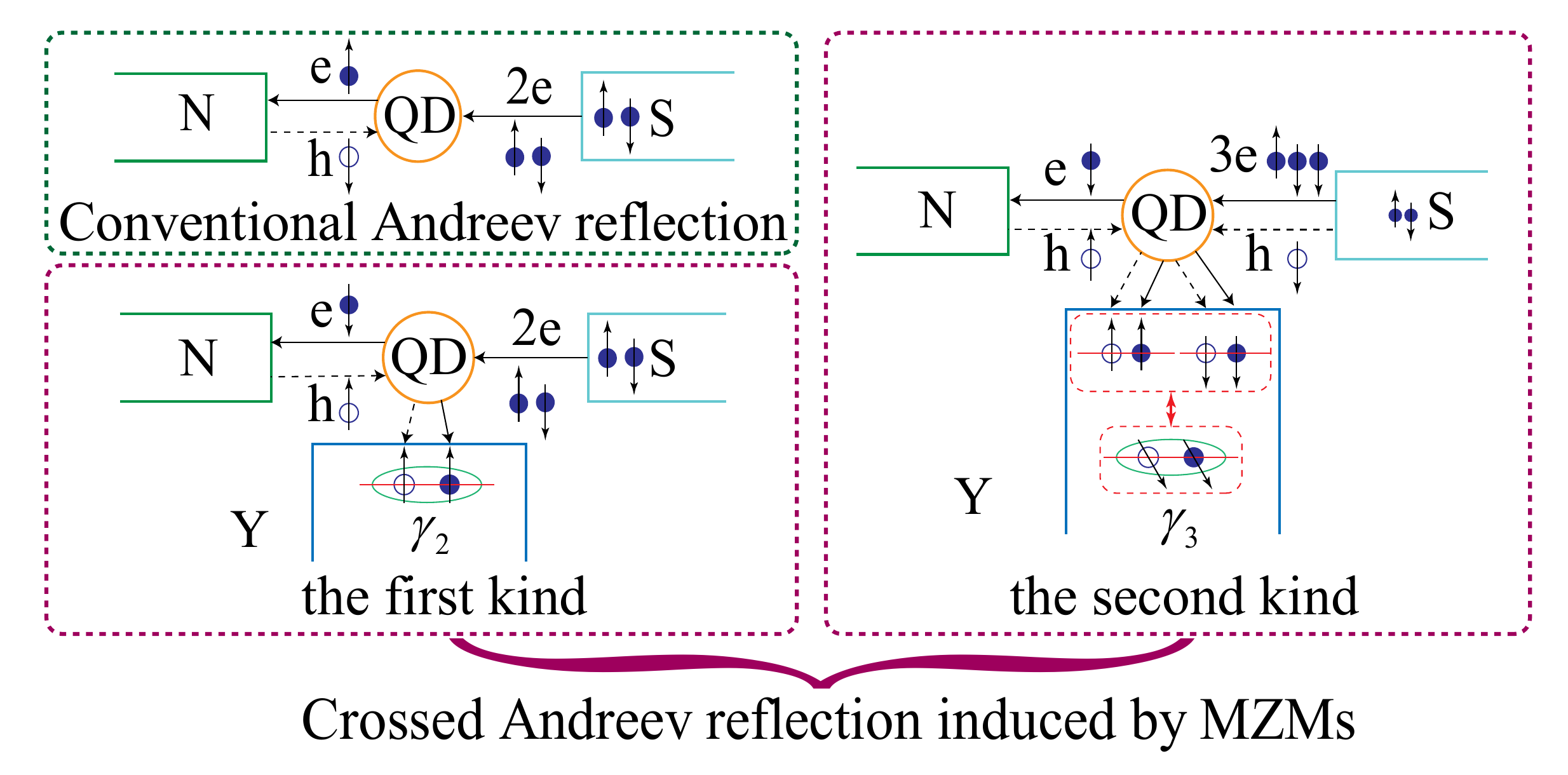}

\caption{\label{fig:Andreev reflection}When the Majorana lead is disconnected
to the QD ($\lambda=0$), electrons in the S lead are transferred
to the N lead through conventional Andreev reflection; when the Majorana
lead is connected to the QD ($\lambda\protect\neq$0), the MZM $\gamma_{2}$
is the coherent superposition of electrons and holes with only spin
$\uparrow$, which leads to the first kind of crossed Andreev reflection.
After braiding, the MZM $\gamma_{3}$ is coupled to the QD. Since
$\gamma_{3}$ is the coherent superposition of electrons and holes
with spins $\uparrow$ and $\downarrow$, the second kind of crossed
Andreev reflection occurs, which is equivalent to the splitting of
charge quanta $3e$. We stress that both kinds of crossed Andreev
reflection exist simultaneously after braiding.}
\end{figure}

\begin{figure}
\includegraphics[width=1\columnwidth]{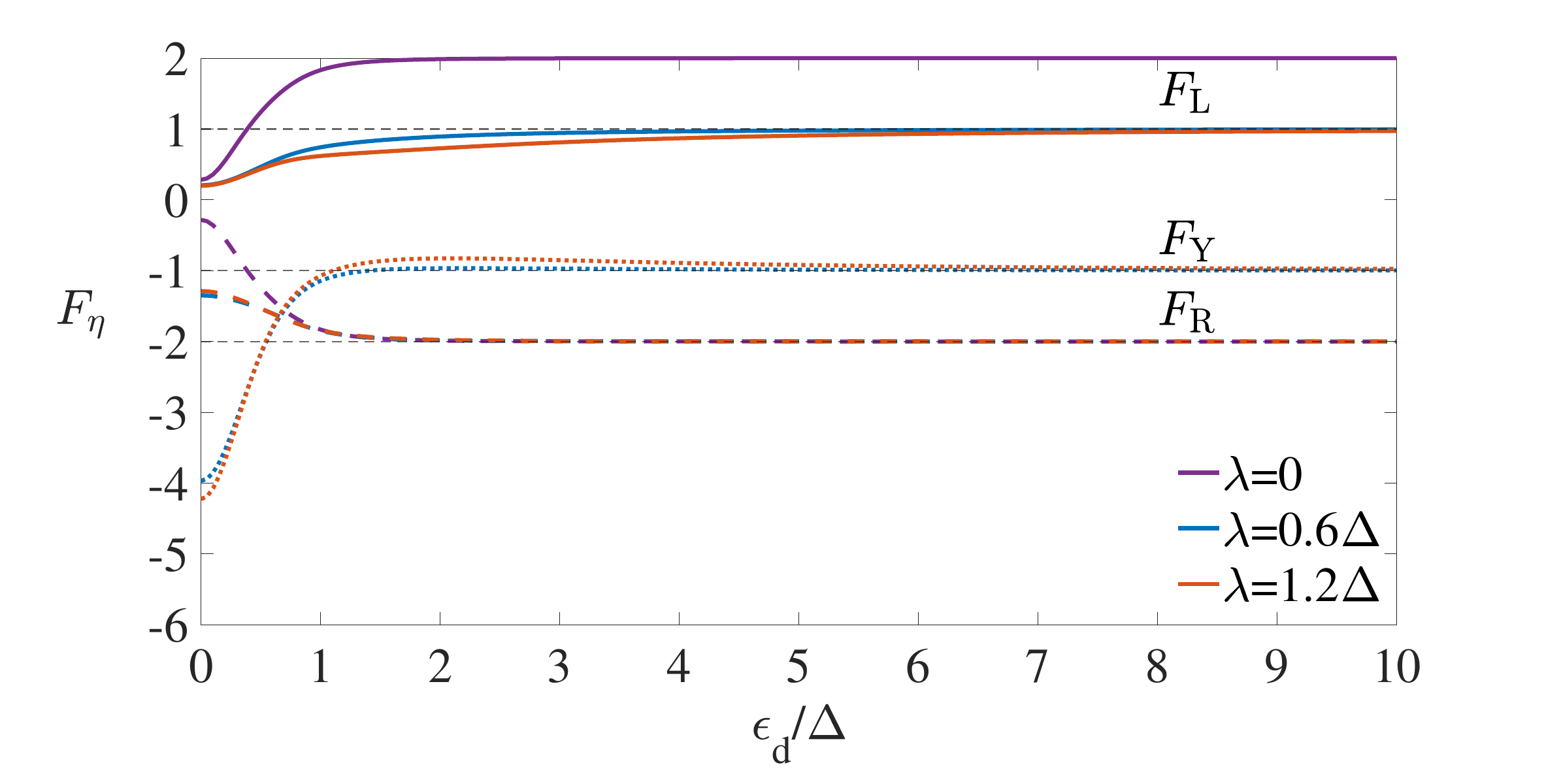}

\caption{\label{fig:Fano-facor}Fano factors at zero temperature of the left
lead (solid lines), right lead (dashed lines) and Majorana lead (dotted
lines) as functions of $\epsilon_{\mathrm{d}}$. The parameters are
$eV_{\mathrm{L}}=0.5\Delta$, $\Gamma_{\mathrm{L}}=\Gamma_{\mathrm{R}}=0.8\Delta$,
$t_{\min}=0.001\Delta$, and $t_{\max}=\Delta$.}

\end{figure}

The discussion will focus on the case of small transparency $(\epsilon_{\mathrm{d}}\gg\Delta$)
since we find that the Fano factors are quantized in this regime.
In Fig.\,\ref{fig:Fano-facor}, we present the Fano factors at zero
temperature as functions of $\epsilon_{\mathrm{d}}/\Delta$ for a
specific realization. In the case of $\lambda=0$, the remaining N-QD-S
junction shows a doubled shot noise in Fano factors $F_{\mathrm{L}}(\epsilon_{\mathrm{d}}\gg\Delta)=-F_{\mathrm{R}}(\epsilon_{\mathrm{d}}\gg\Delta)=2$
due to the transport of Cooper pairs through conventional Andreev
reflection \citep{PhysRevB.49.16070}. When the Majorana Y junction
is connected to the QD with $\lambda\neq0$, we find $F_{\mathrm{L}}(\epsilon_{\mathrm{d}}\gg\Delta)=-F_{\mathrm{Y}}(\epsilon_{\mathrm{d}}\gg\Delta)=1$
and $|F_{\mathrm{R}}(\epsilon_{\mathrm{d}}\gg\Delta)|=2$. The results
denote that the unit of charge transferred between the QD and the
normal-metal lead is $e$ as well as the Majorana lead, while the
unit of charge transferred between the QD and the superconducting
lead is $2e$. This is the process of the first kind of crossed Andreev
reflection. A hole from the left lead is reflected as an electron
into the Majorana lead, while a Cooper pair from the superconducting
lead is reflected as a hole into the Majorana lead and an electron
into the normal-metal lead \citep{PhysRevLett.101.120403}, as shown
in Fig.\,\ref{fig:Andreev reflection}. The holes transferred through
crossed Andreev reflection act as facilitators to propel the splitting
of Cooper pairs, and do not contribute to the transferred charge.
In this regime, local Andreev reflection is fully suppressed, and
the first kind of crossed Andreev reflection dominates.

\textcolor{black}{The quantized Fano factors result from the occurrence
of the YSR state below the gap. As shown in Fig.\,\ref{fig:LDOS 1},
a very sharp Majorana-induced mid-gap resonance at $\omega=0$ when
$\epsilon_{\mathrm{d}}\gg\Delta$ is regarded as a YSR bound state.
The occurrence of the YSR state means that the Cooper pairing is ``softened''
so that the free electrons from the superconducting lead screen the
spins of the QD \citep{HEINRICH20181,2013PhyOJ...6...75M}. The state-screened
process leads to the splitting of Cooper pairs. In this way, crossed
Andreev reflection dominates the tunneling processes, which gives
rise to the quantized Fano factor.}

\section{SIGNATURES OF THE MAJORANA BRAIDING}

Now we braid the MZMs by taking $\gamma_{2}\rightarrow-\gamma_{3}$
and $\gamma_{3}\rightarrow\gamma_{2}$. Since the spin orientations
of the MZMs $\gamma_{2}$ and $\gamma_{3}$ belonging to the same
complex fermion $c_{2}$ are different, observable consequences can
be obtained with the connection to the QD. The QD is connected to
the Majorana Y junction through $\gamma_{3}$. Using the Nambu spinors
and Eq.\,\eqref{eq:Hamiltonian of the Y junction}, we can easily
obtain the Hamiltonian of the Majorana Y junction $\widetilde{\mathcal{H}}_{\mathrm{Y}}$
after braiding. Given that the angle of spin orientations between
$\gamma_{2}$ and $\gamma_{3}$ is $\theta$, the spin-conserving
coupling is then given by $\widetilde{H}_{\mathrm{T,Y}}=-\lambda\widetilde{d}_{\uparrow}\gamma_{3}+\mathrm{H.c.}$,
where $\widetilde{d}_{\uparrow}$ $(\widetilde{d}_{\downarrow})$
is the electron operators of the QD with the same (opposite) spin
orientation described by $\gamma_{3}$ with

\begin{equation}
\left(\begin{array}{c}
\widetilde{d}_{\uparrow}\\
\widetilde{d}_{\downarrow}
\end{array}\right)=\left(\begin{array}{cc}
\cos\frac{\theta}{2} & \sin\frac{\theta}{2}\\
-\sin\frac{\theta}{2} & \cos\frac{\theta}{2}
\end{array}\right)\left(\begin{array}{c}
d_{\uparrow}\\
d_{\downarrow}
\end{array}\right).\label{eq:so(2) rotation}
\end{equation}

\begin{figure*}
\subfloat[\label{fig:Conductance of the left lead after braiding 1}]{\includegraphics[width=0.65\columnwidth]{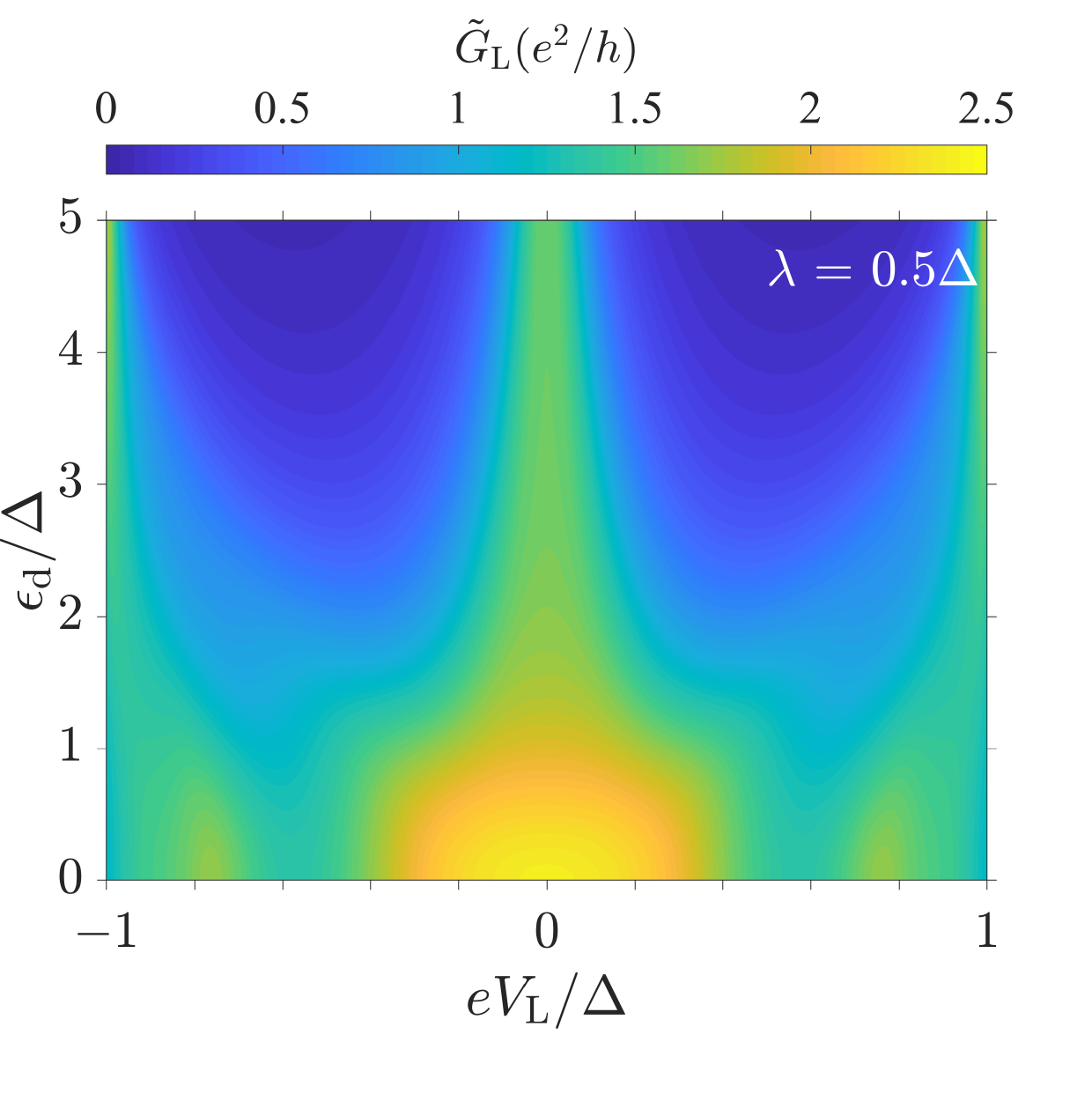}}
\subfloat[\label{fig::Conductance of the left lead after braiding 2}]{\includegraphics[width=0.65\columnwidth]{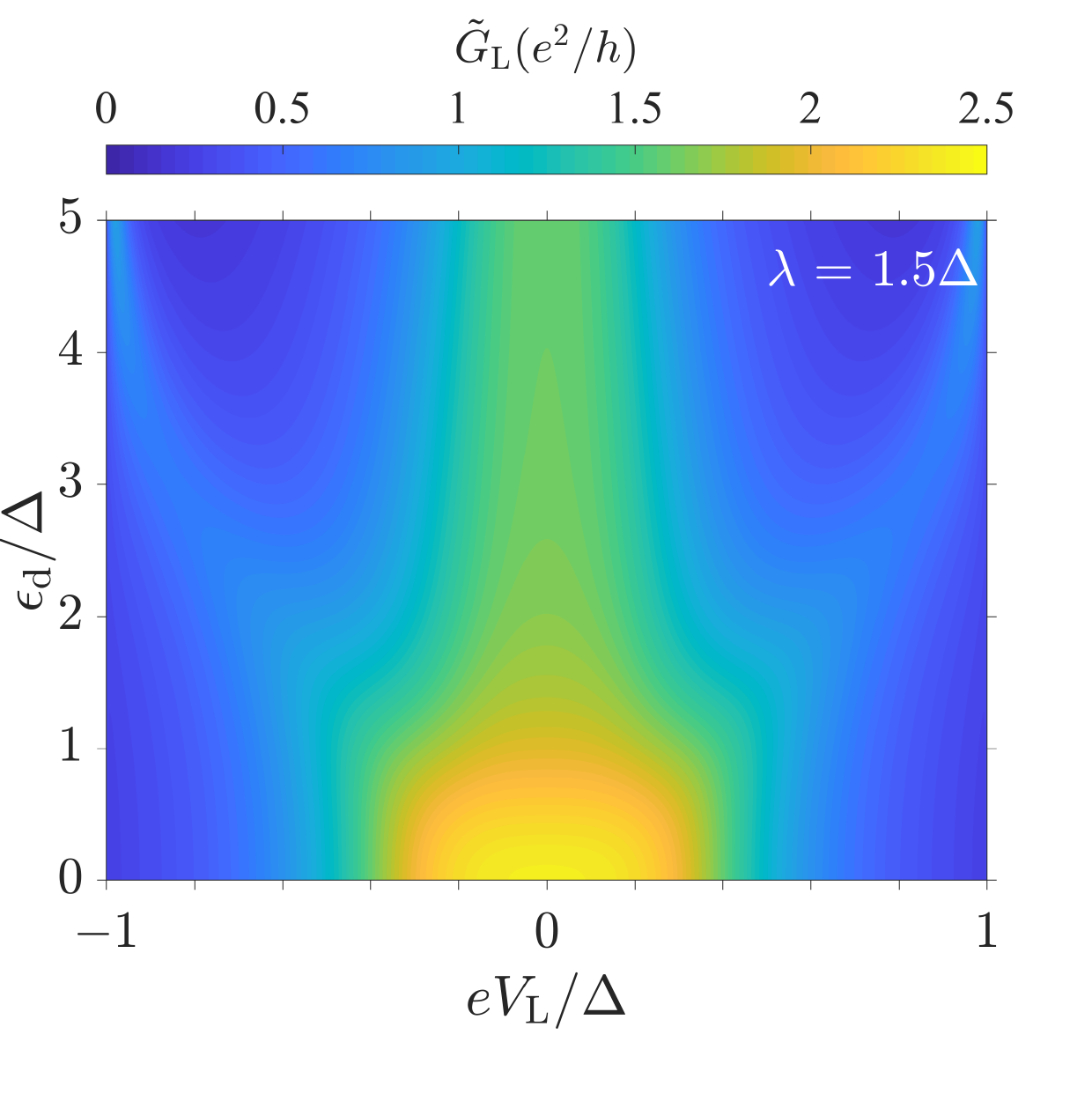}

}\subfloat[\label{fig::Conductance of the left lead after braiding 3}]{\includegraphics[width=0.65\columnwidth]{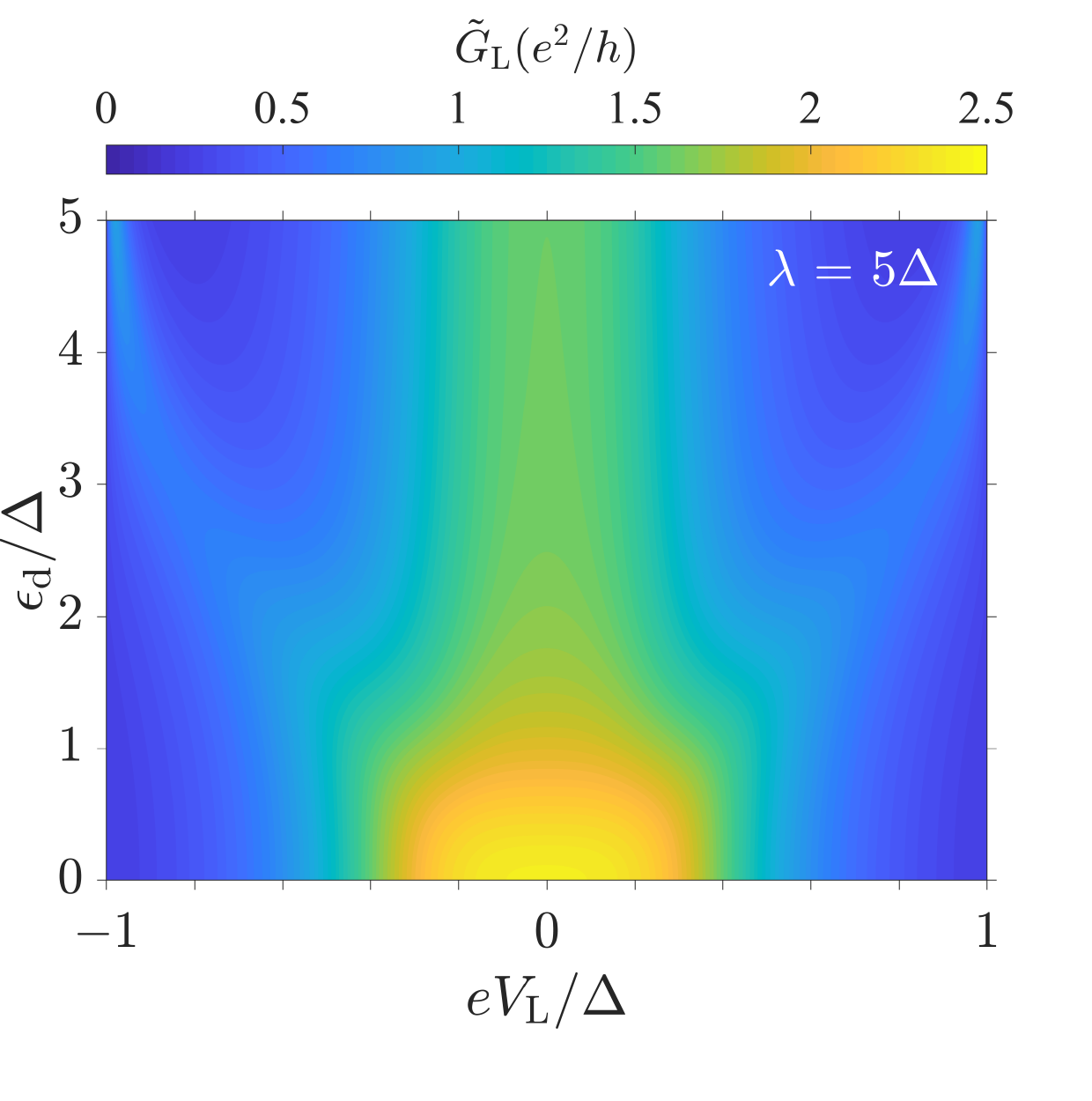}

}\caption{\label{fig:The differential conductances after braiding}Differential
conductance spectra of the normal-metal lead as a function of $eV_{\mathrm{L}}/\Delta$
and $\epsilon_{\mathrm{d}}/\Delta$ after braiding at zero temperature.
The parameters are the same as in Fig.\,\ref{fig:The differential conductances}}
\end{figure*}

For the Majorana Y junction sketched in Fig.\,\ref{fig: Majorana-Y-junction},
the spin orientation angle is $\theta=\frac{2}{3}\pi$. Such a braiding
process is equivalent to involving spin-flip tunneling between the
QD and Majorana lead. After braiding, we can obtain the ZBCPs by

\begin{align}
\lim_{eV_{\mathrm{L}}\to0}\widetilde{G}_{\mathrm{L}}(eV_{\mathrm{L}}) & =\begin{cases}
\frac{e^{2}}{h}\frac{8\Gamma_{\mathrm{L}}^{2}((\ensuremath{\Gamma_{\mathrm{L}}^{2}}+\ensuremath{\Gamma_{\mathrm{R}}^{2}}+4\ensuremath{\epsilon_{\mathrm{d}}^{2}})^{2}+\Gamma_{\mathrm{R}}^{4}+\Gamma_{\mathrm{L}}^{2}\Gamma_{\mathrm{R}}^{2})}{(4\Gamma_{\mathrm{L}}^{2}+\Gamma_{\mathrm{R}}^{2})(\Gamma_{\mathrm{L}}^{2}+\Gamma_{\mathrm{R}}^{2}+4\epsilon_{\mathrm{d}}^{2})^{2}},\,\lambda\neq0,\\
\frac{e^{2}}{h}\frac{16\Gamma_{\mathrm{L}}^{2}\Gamma_{\mathrm{R}}^{2}}{(\Gamma_{\mathrm{L}}^{2}+\Gamma_{\mathrm{R}}^{2}+\epsilon_{\mathrm{d}}^{2})^{2}},\,\thinspace\thinspace\thinspace\thinspace\thinspace\thinspace\thinspace\thinspace\thinspace\thinspace\thinspace\thinspace\thinspace\thinspace\thinspace\thinspace\thinspace\thinspace\thinspace\thinspace\thinspace\thinspace\thinspace\thinspace\thinspace\thinspace\thinspace\thinspace\thinspace\thinspace\thinspace\lambda=0,
\end{cases}\label{eq:zero-bias conductance of the left lead after braiding}\\
\lim_{eV_{\mathrm{L}}\to0}\widetilde{G}_{\mathrm{R}}(eV_{\mathrm{L}}) & =\begin{cases}
-\frac{e^{2}}{h}\frac{4\Gamma_{\mathrm{L}}^{2}\Gamma_{\mathrm{R}}^{2}(5\Gamma_{\mathrm{L}}^{2}+5\Gamma_{\mathrm{R}}^{2}+12\epsilon_{\mathrm{d}}^{2})}{\text{(4\ensuremath{\Gamma_{\mathrm{L}}^{2}}+\ensuremath{\Gamma_{\mathrm{R}}^{2}})(\ensuremath{\Gamma_{\mathrm{L}}^{2}}+\ensuremath{\Gamma_{\mathrm{R}}^{2}}+4\ensuremath{\epsilon_{\mathrm{d}}^{2}})}^{2}}, & \lambda\neq0,\\
-\frac{e^{2}}{h}\frac{16\Gamma_{\mathrm{L}}^{2}\Gamma_{\mathrm{R}}^{2}}{(\Gamma_{\mathrm{L}}^{2}+\Gamma_{\mathrm{R}}^{2}+\epsilon_{\mathrm{d}}^{2})^{2}}, & \lambda=0,
\end{cases}\\
\lim_{eV_{\mathrm{L}}\to0}\widetilde{G}_{\mathrm{Y}}(eV_{\mathrm{L}}) & =\begin{cases}
-\frac{e^{2}}{h}\frac{4\Gamma_{\mathrm{L}}^{2}(2\Gamma_{\mathrm{L}}^{2}-\Gamma_{\mathrm{R}}^{2}+8\epsilon_{\mathrm{d}}^{2})}{\text{(4\ensuremath{\Gamma_{\mathrm{L}}^{2}}+\ensuremath{\Gamma_{\mathrm{R}}^{2}})(\ensuremath{\Gamma_{\mathrm{L}}^{2}}+\ensuremath{\Gamma_{\mathrm{R}}^{2}}+4\ensuremath{\epsilon_{\mathrm{d}}^{2}})}}\text{,} & \lambda\neq0\text{,}\\
0\text{,} & \lambda=0.
\end{cases}
\end{align}

\begin{figure}[H]
\includegraphics[width=1\columnwidth]{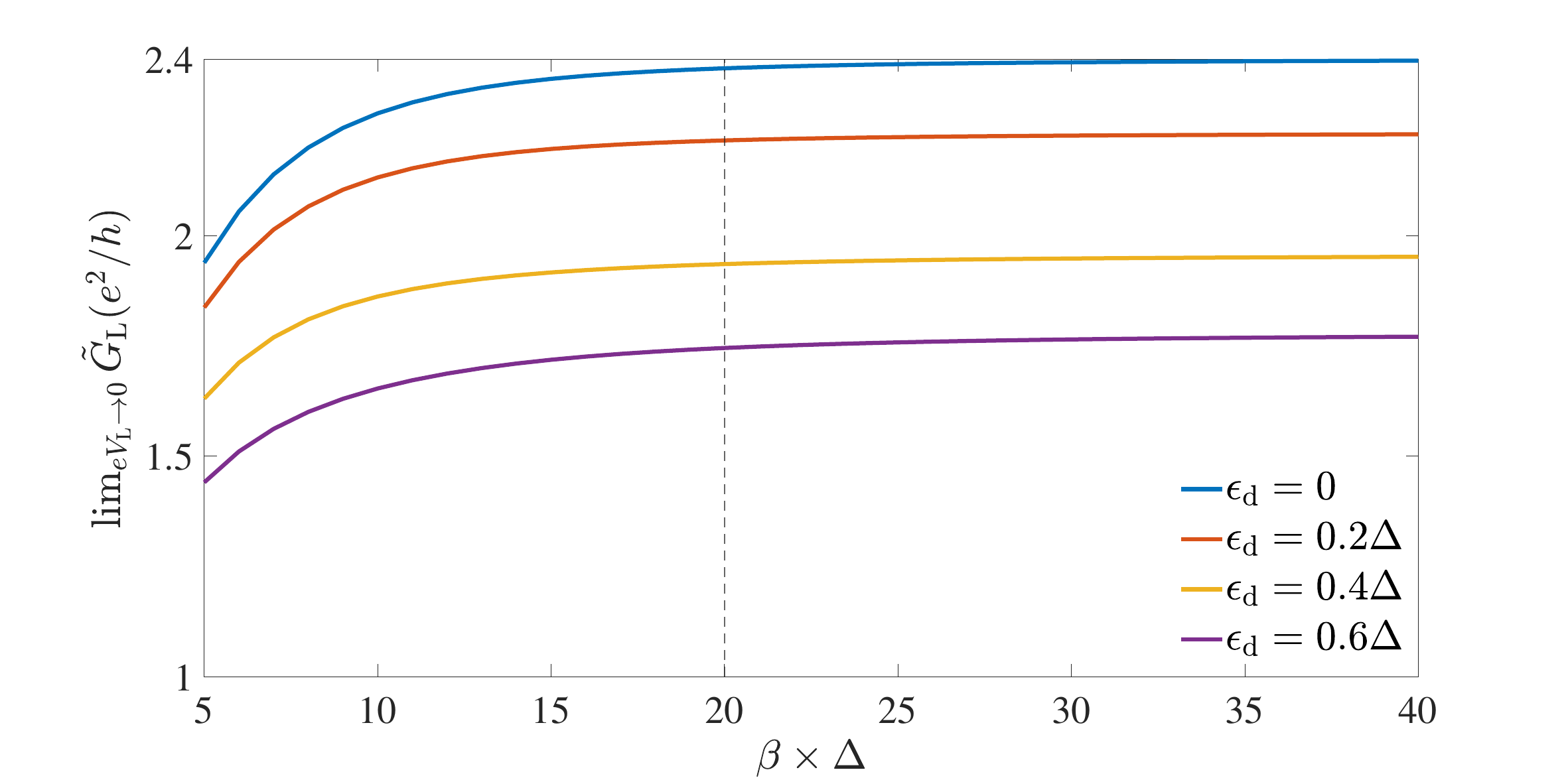}

\caption{\label{fig:Plot of peak vs temperature after braiding} ZBCP of the
normal-metal lead after braiding as a function of the finite temperature.
The parameters are the same as in Fig.\,\ref{fig: Plot of peak vs temperature}.}
\end{figure}

When $\lambda=0$, the result is the same as that before braiding;
when $\lambda\neq0$, the occurrence of spin-flip tunneling shifts
the ZBCP. We plot the differential conductance of the normal-metal
lead after the Majorana braiding in Fig.\,\ref{fig:The differential conductances after braiding}
for comparison to Fig.\,\ref{fig:The differential conductances}.
Particularly, if the QD is symmetrically coupled ($\Gamma_{\mathrm{L}}=\Gamma_{\mathrm{R}}=\Gamma$)
, the ZBCP of the normal-metal lead maximally shifts to $2.4e^{2}/h$
for $\epsilon_{\mathrm{d}}=0$ and $1.6e^{2}/h$ for $\epsilon_{\mathrm{d}}\gg\Delta$,
which can act a robust hint of Majorana braiding. With the increasing
of $\epsilon_{\mathrm{d}}$, the ZBCP gets broadened and its height
gets lower concurrently. As shown in Fig.\,\ref{fig:Plot of peak vs temperature after braiding}.,
the thermal broadening effect can also be suppressed by taking $\epsilon_{\mathrm{d}}=0$
and each solid line is closed to the corresponding zero-temperature
limit when $k_{B}T<\Delta/20$. Consequently, it is appropriate to
observe the ZBCP with large junction transparency.

\begin{figure}
\includegraphics[width=1\columnwidth]{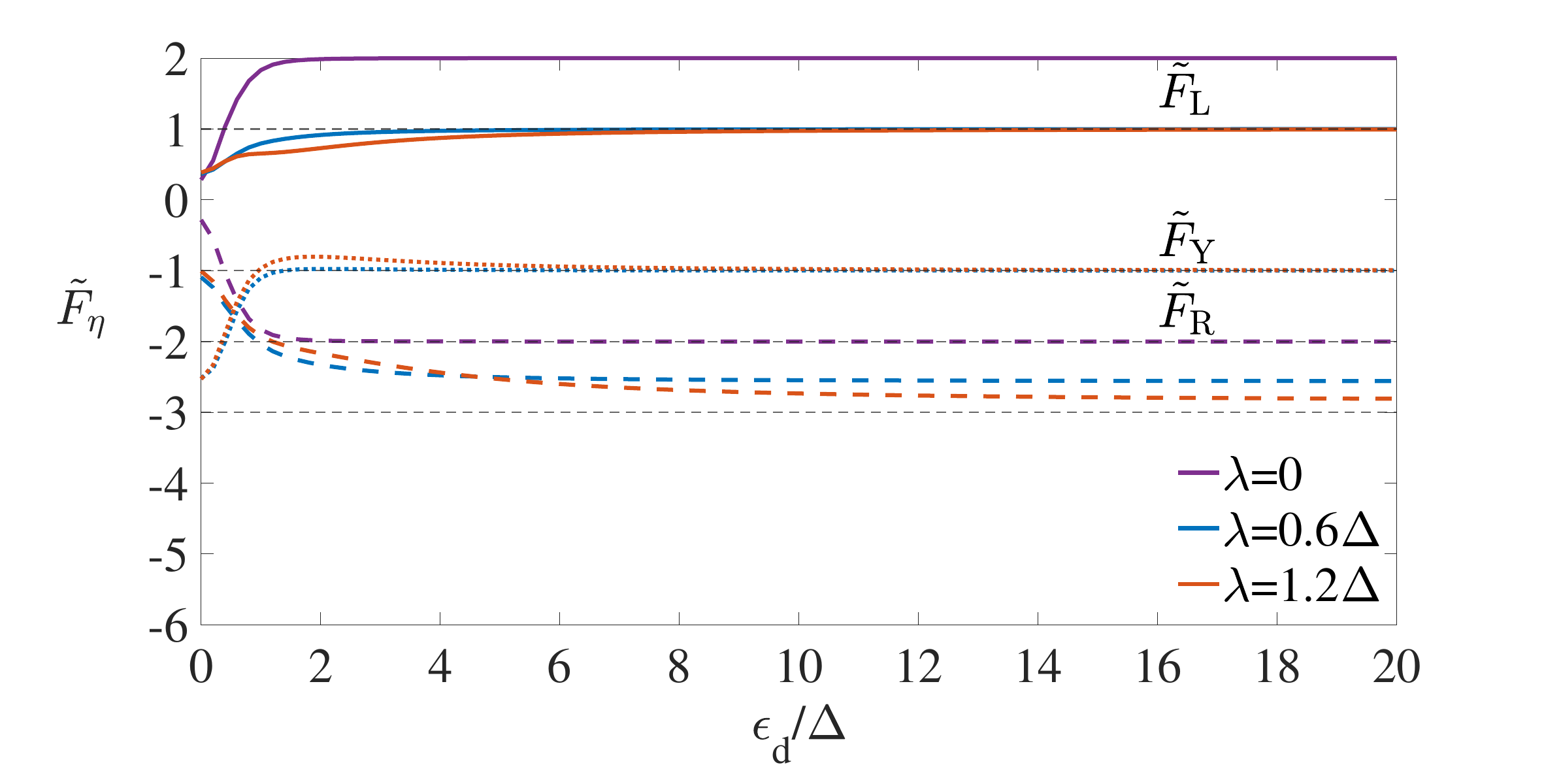}

\caption{\label{fig:Fano factor after braiding}Fano factors after braiding
at zero temperature of the left lead (solid lines), right lead (dashed
lines) and Majorana lead (dotted lines) as functions of $\epsilon_{\mathrm{d}}$.
The parameters are the same as in Fig.\,\ref{fig:Fano-facor}.}
\end{figure}

The Fano factors after braiding are also quantized, but $2<|\widetilde{F}_{\mathrm{R}}(\epsilon_{\mathrm{d}}\gg\Delta)|<3$,
\emph{i.e}., the unit of charge transferred between the QD and the
superconducting lead is larger than that of a Cooper pair, as shown
Fig.\,\ref{fig:Fano factor after braiding}. This result is induced
by involving both spins $\uparrow$ and $\downarrow$ in the coupling
between the QD and the MZM $\gamma_{3}$. As illustrated in Fig.\,\ref{fig:Andreev reflection},
the second kind of crossed Andreev reflection occurs after the Majorana
braiding.

Specifically, a Cooper pair transferred between the QD and the superconducting
lead is accompanied by an extra electron and a hole, which leads to
the $3e$ charge quanta. One electron of the $3e$ charge quanta is
reflected as an electron into the normal-metal lead, while the other
two are reflected as holes into the Majorana lead. Such a process
of charge transmission is equivalent to the splitting of the $3e$
charge quanta.

Given that the electrons coupled to the MZM $\gamma_{3}$ are composed
of spin-$\uparrow$ and $\downarrow$ electrons with a certain weight
depending on the angle $\theta$ (see Eq.\,\eqref{eq:so(2) rotation}),
both kinds of crossed Andreev reflection exist simultaneously, which
leads to $2<|\widetilde{F}_{\mathrm{R}}(\epsilon_{\mathrm{d}}\gg\Delta)|<3$.
As shown in Fig.\,\ref{fig:Fano factor after braiding}, the second
kind of crossed Andreev reflection gains the dominance of the tunneling
processes (corresponding to $|\widetilde{F}_{\mathrm{R}}(\epsilon_{\mathrm{d}}\gg\Delta)|\rightarrow3$)
with increasing $\lambda$. As for the Majorana lead, the acceptance
of spin-$\uparrow$ and $\downarrow$ electrons with a certain weight
is equivalent to the acceptance of an electron with spin polarization
angle $\theta$ in each current pulse, which gives rise to $|\widetilde{F}_{\mathrm{Y}}(\epsilon_{\mathrm{d}}\gg\Delta)|=1$.
The units of the charge transferred between the normal-metal lead
and the QD for both kinds of crossed Andreev reflection are identical,
so the Fano factor of the normal-metal lead stays at $|\widetilde{F}_{\mathrm{L}}(\epsilon_{\mathrm{d}}\gg\Delta)|=1$,
the same as that before braiding.

\section{CONCLUSION}

We have studied the ZBCPs and the Fano factors of the T-shaped structure.
We have shown that the ZBCP of the normal-metal lead is always quantized
to $2e^{2}/h$ at zero temperature before braiding, which is quite
robust at finite temperature when the QD is on-resonance. This quantized
conductance can entirely arise from the Majorana-induced crossed Andreev
reflection, which is protected by the energy gap of the superconducting
lead. After Majorana braiding, the quantized ZBCP shifts and becomes
dependent on the line widths $\Gamma_{\mathrm{L}}$, $\Gamma_{\mathrm{R}}$
and the QD level $\epsilon_{\mathrm{d}}$. This variation is owing
to the introduction of spin-flip tunneling between the Majorana lead
and the QD after braiding. By analyzing the quantized Fano factors,
we have found that the crossed Andreev reflection dominates over the
conventional Andreev reflection when $\epsilon_{\mathrm{d}}\gg\Delta$.
We have also found a novel kind of crossed Andreev reflection equivalent
to the splitting of the $3e$ charge quanta. The quantized ZBCPs and
Fano factors induced by the nonlocal crossed Andreev reflection provide
strong fingerprint for MZMs.

\appendix

\section{CALCULATION OF $\Sigma_{\mathrm{Y}}^{<}$\label{sec:CALCULATION OF THE SELF-ENERGY}}

In this section, we present the details of the analytical calculation
of the terms containing $\Sigma_{\mathrm{Y}}^{<}$ in Eq.\,\ref{eq:Current in GF form}
. The lesser self-energy from the Majorana lead is given by
\begin{equation}
\Sigma_{\mathrm{Y}}^{<}=F_{\mathrm{Y}}(\Sigma_{\mathrm{Y}}^{A}-\Sigma_{\mathrm{Y}}^{R})=-i2F_{\mathrm{Y}}\mathrm{Im}\Sigma_{\mathrm{Y}}^{R},
\end{equation}
where $\Sigma_{\mathrm{Y}}^{R}$ is determined by
\begin{align}
\kappa & \thickapprox\frac{-8\lambda^{2}t_{\max}^{2}}{-4t_{\mathrm{\max}}^{2}\omega^{+}+(\omega^{+})^{3}}+\frac{2\lambda^{2}(\omega^{+})^{2}}{-4t_{\mathrm{\max}}^{2}\omega^{+}+(\omega^{+})^{3}},
\end{align}
with $\omega^{+}=\omega+i0^{+}$, as we mentioned in the main text
($\kappa$ here is same to that appeared in Eq.\,\eqref{eq:self-energy of the Majorana lead}).
We use the notation $\kappa_{1}$and $\kappa_{2}$ to denote the first
term and the second term of $\kappa$, respectively. Neglecting the
high-order terms of $(i0^{+})^{n\ge2}$, we can obtain the following
expressions:

\begin{align}
\kappa_{1} & =\frac{-8\lambda^{2}t_{\max}^{2}}{(-4t_{\mathrm{\max}}^{2}\omega+\omega^{3})-(4t_{\mathrm{\max}}^{2}-3\omega^{2})(i0^{+})},\\
\kappa_{2} & =\frac{2\lambda^{2}\omega^{2}}{\omega(-4t_{\mathrm{\max}}^{2}+\omega^{2})+(4t_{\mathrm{\max}}^{2}+\omega^{2})(i0^{+})}.
\end{align}
Using the formula
\begin{equation}
\lim_{\eta\rightarrow0^{+}}\frac{1}{x\pm i\eta}=\mathcal{P}\frac{1}{x}\mp i\pi\delta(x),
\end{equation}
we can obtain
\begin{align}
i\mathrm{Im(\kappa_{1})} & =i\pi\frac{-8\lambda^{2}t_{\max}^{2}}{4t_{\max}^{2}-3\omega^{2}}\delta\left(\frac{-4t_{\max}^{2}\omega+\omega^{3}}{4t_{\max}^{2}-3\omega^{2}}\right),\\
i\mathrm{Im}(\kappa_{2}) & =-i\pi\frac{2\lambda^{2}\omega^{2}}{4t_{\max}^{2}+\omega^{2}}\delta\left(\frac{-4t_{\max}^{2}\omega+\omega^{3}}{4t_{\max}^{2}+\omega^{2}}\right).
\end{align}
Using the relationship

\begin{equation}
\delta(\phi(x))=\sum_{j}\frac{1}{|\phi'(x)|}\delta(x-x_{j}),
\end{equation}
with $\phi(x_{j})=0$, the imaginary part of $\kappa_{1}$and $\kappa_{2}$
reduce to
\begin{align}
i\mathrm{Im(\kappa_{1})} & =Q_{1}(\omega)[\delta(\omega)+\delta(\omega-2t_{\max})+\delta(\omega+2t_{\max})],\\
i\mathrm{Im(\kappa_{2})} & =Q_{2}(\omega)[\delta(\omega)+\delta(\omega-2t_{\max})+\delta(\omega+2t_{\max})],
\end{align}

\begin{widetext}
where

\begin{align}
Q_{1}(\omega) & =i\pi\frac{-8\lambda^{2}t_{\max}^{2}}{4t_{\max}^{2}-3\omega^{2}}\left|\frac{(4t_{\max}^{2}-3\omega^{2})^{2}}{-(4t_{\max}^{2}-3\omega^{2})^{2}-6\omega^{2}(4t_{\max}^{2}-\omega^{2})}\right|,\\
Q_{2}(\omega) & =-i\pi\frac{2\lambda^{2}\omega^{2}}{4t_{\max}^{2}+\omega^{2}}\left|\frac{(4t_{\max}^{2}+\omega^{2})^{2}}{(-4t_{\max}^{2}+3\omega^{2})(4t_{\max}^{2}+\omega^{2})-2\omega^{2}(-4t_{\max}^{2}+\omega^{2})}\right|.
\end{align}
\end{widetext}

It is obvious that $Q_{1}(0)=-2i\pi\lambda^{2}$, $Q_{1}(\pm2t_{\max})=i\pi\lambda^{2}$,
$Q_{2}(0)=0$ and $Q_{2}(\pm2t_{\max})=-i\pi\lambda^{2}$. Since the
electron-hole symmetry gives $G_{\mathrm{QD}}^{R}(-\omega)=-\left[G_{\mathrm{QD}}^{R}(\omega)\right]^{*}$,
\emph{i.e.}, $G_{\mathrm{QD}}^{R}(0)$ is a purely imaginary function.
Note that the imaginary part of $G_{\mathrm{QD}}^{R}(0)$ is tiny,
we can obtain $G_{\mathrm{QD}}^{R}(0)\approx0$. For example, the
terms containing $\Sigma_{\mathrm{Y}}^{<}$ in Eq.\,\eqref{eq:Current in GF form}
are calculated by

\begin{align}
\int\mathrm{d}\omega G_{\mathrm{QD}}^{R}(\omega)\Sigma_{\mathrm{Y}}^{<}(\omega) & \propto\int\mathrm{d}\omega G_{\mathrm{QD}}^{R}(\omega)[i\mathrm{Im}(\kappa_{1}+\kappa_{2})]\nonumber \\
 & =-2i\pi\lambda^{2}G_{\mathrm{QD}}^{R}(0)\nonumber \\
 & =0,
\end{align}

and

\begin{align}
 & \int\mathrm{d}\omega G_{\mathrm{QD}}^{R}(\omega)\Sigma_{\mathrm{Y}}^{<}(\omega)G_{\mathrm{QD}}^{A}(\omega)\Sigma_{\mathrm{\eta}}^{A}(\omega)\nonumber \\
 & =i\pi\lambda^{2}\left[G_{\mathrm{QD}}^{R}(0)\Lambda G_{\mathrm{QD}}^{A}(0)\Sigma_{\mathrm{\eta}}^{A}(0)\right]\nonumber \\
 & =0.
\end{align}
Hence we can take $\Sigma_{\mathrm{Y}}^{<}=0$ in the calculation
of the time-average current and the shot noise.

\section{CALCULATION OF THE LDOS OF THE SUPERCONDUCTING LEAD\label{sec:CALCULATION-OF-LDOS}}
\begin{widetext}
For convenience, the whole system is divided into two subsystems,
one is \textquotedbl quantum dot + Majorana Y-junction+Normal-metal
lead\textquotedbl , the other is the superconducting lead. The Hamiltonian
of the superconducting lead is given by Eq.\,(3) in the main text,

\[
H_{\mathrm{R}}=\sum_{k\sigma}\epsilon_{\mathrm{R},k\sigma}a_{\mathrm{R},k\sigma}^{\dagger}a_{\mathrm{R},k\sigma}+\sum_{k}(\Delta a_{\mathrm{R},k\uparrow}^{\dagger}a_{\mathrm{R},-k\downarrow}^{\dagger}+\mathrm{H.c.}).
\]
In the Nambu space $(a_{\mathrm{R},k\uparrow}^{\dagger},a_{\mathrm{R},-k\downarrow},a_{\mathrm{R},-k\downarrow}^{\dagger},a_{\mathrm{R},k\uparrow})$,
the unperturbed Green's function of the BCS superconductor evaluated
at the origin ($r=0$) is represented as

\begin{align*}
g_{\mathrm{R}}^{R}(\omega) & =\int\mathrm{d}^{3}k\frac{e^{ik\cdot r}}{\omega_{+}^{2}-\xi_{k}^{2}-\Delta^{2}}\left(\begin{array}{cccc}
\omega_{+}+\xi_{k} & \Delta\\
\Delta & \omega_{+}-\xi_{k}\\
 &  & \omega_{+}+\xi_{k} & \Delta\\
 &  & \Delta & \omega_{+}-\xi_{k}
\end{array}\right)\\
 & =-\rho_{R}\int\mathrm{d}\xi_{k}\frac{1}{\left(\xi_{k}-\sqrt{\omega_{+}^{2}-\Delta^{2}}\right)\left(\xi_{k}+\sqrt{\omega_{+}^{2}-\Delta^{2}}\right)}\left(\begin{array}{cccc}
\omega_{+}+\xi_{k} & \Delta\\
\Delta & \omega_{+}-\xi_{k}\\
 &  & \omega_{+}+\xi_{k} & -\Delta\\
 &  & -\Delta & \omega_{+}-\xi_{k}
\end{array}\right),
\end{align*}
where $\rho_{R}$ is the density of states and $\omega_{+}=\omega+i\eta$
with $\eta=0^{+}$. We reduce the expression above by taking

\[
\sqrt{\omega_{+}^{2}-\Delta^{2}}=\sqrt{\omega^{2}+i\omega\eta-\Delta^{2}}\approx\sqrt{\omega^{2}-\Delta^{2}}+i\mathrm{sgn}(\omega)\eta.
\]
When $|\Delta|>|\omega|$, the pole (imaginary part >0) is $\xi_{k}=$$i\sqrt{\Delta^{2}-\omega{}^{2}}$;
when $|\Delta|<|\omega|$, the pole (imag part >0) is $\xi_{k}=i\eta$+$sgn(\omega)\sqrt{\omega{}^{2}-\Delta^{2}}$.
Using the theorem of residues, we obtain the unperturbed Green's function
of the BCS superconductor as

\begin{align*}
g_{\mathrm{R}}^{R}(\omega) & =\left[-\theta(|\Delta|-|\omega|)\pi\rho_{R}\frac{1}{\sqrt{\Delta^{2}-\omega{}^{2}}}-i\theta(|\omega|-|\Delta|)\pi\rho_{R}\frac{sgn(\omega)}{\sqrt{\omega{}^{2}-\Delta^{2}}}\right]\left(\begin{array}{cccc}
\omega & \Delta\\
\Delta & \omega\\
 &  & \omega & -\Delta\\
 &  & -\Delta & \omega
\end{array}\right)\\
 & =-i\pi\rho_{R}\beta\left(\begin{array}{cccc}
1 & \frac{\Delta}{\omega}\\
\frac{\Delta}{\omega} & 1\\
 &  & 1 & -\frac{\Delta}{\omega}\\
 &  & -\frac{\Delta}{\omega} & 1
\end{array}\right),
\end{align*}
where $\beta(\omega)=\theta(|\Delta|-|\omega|)\frac{\omega}{i\sqrt{\Delta^{2}-\omega{}^{2}}}+\theta(|\omega|-|\Delta|)\frac{|\omega|}{\sqrt{\omega{}^{2}-\Delta^{2}}}$.
The spectral function of the superconducting lead is given by

\[
A_{\mathrm{R}}(\omega)=i(G_{\mathrm{R}}^{R}(\omega)-G_{\mathrm{R}}^{A}(\omega))=-2\mathrm{Im}G_{\mathrm{R}}^{R}(\omega)
\]
with

\[
G_{\mathrm{R}}^{R}(\omega)=\left(\left(g_{\mathrm{R}}^{R}(\omega)\right)^{-1}-\Sigma_{\mathrm{QD}}^{R}(\omega)\right)^{-1}.
\]
The self-energy at the origin ($r=0$) can be calculated as
\begin{align*}
\Sigma_{\mathrm{QD}}^{R}(\omega) & =\sum_{k}\mathcal{H}_{\mathrm{T,R}}^{\dagger}G_{\mathrm{QD}}^{R}(\omega)\mathcal{H}_{\mathrm{T,R}}\\
 & =\sum_{k}|\nu_{\mathrm{R,k}}|^{2}G_{\mathrm{QD}}^{R}(\omega)\\
 & =2\pi\rho_{\mathrm{R}}|\nu_{\mathrm{R,r=0}}|^{2}G_{\mathrm{QD}}^{R}(\omega)\\
 & =\Gamma_{\mathrm{R}}G_{\mathrm{QD}}^{R}(\omega),
\end{align*}
where we have used Fourier transform $|\nu_{\mathrm{R,r}}|^{2}=\frac{1}{2\pi}\sum_{k}e^{ikr}|\nu_{\mathrm{R,k}}|^{2}$.
The density of states $\rho_{\mathrm{R}}$ around the Fermi-surface
is approximately regarded as a constant. The effective Green's function
of the QD in the subsystem \textquotedbl quantum dot + Majorana Y-junction+Normal-metal
lead\textquotedbl{} is given by
\[
G_{\mathrm{QD}}^{R}(\omega)=\left(\left(g_{\mathrm{QD}}^{R}(\omega)\right)^{-1}-\Sigma_{\mathrm{L}}^{R}(\omega)-\Sigma_{\mathrm{Y}}^{R}(\omega)\right)^{-1}.
\]
Hence the LDOS of the superconducting lead is
\[
\rho(\omega)=\mathrm{Tr}\left[A_{\mathrm{R}}(\omega)\right]/2\pi.
\]
\end{widetext}

\section{CALCULATION OF THE SHOT NOISE $S_{\eta\eta'}(\omega')$\label{sec:CALCULATION OF THE SHOT NOISE}}

In this section,we review the formalism for the shot noise which will
be used in the main text \citep{PhysRevB.78.155303,PhysRevB.85.085415}.
We consider the Hamiltonian of a multi-terminal systems with a noninteracting
central QD
\begin{equation}
H=\sum_{\eta}H_{\eta}+H_{\mathrm{QD}}+H_{T},
\end{equation}
where $H_{\mathrm{\eta}}=\sum_{k\sigma}\epsilon_{\mathrm{\eta},k\sigma}a_{\mathrm{\mathrm{\eta}},k\sigma}^{\dagger}a_{\mathrm{\mathrm{\eta}},k\sigma}$
and $H_{\mathrm{QD}}=\sum_{\sigma}\epsilon_{\mathrm{d}}d_{\sigma}^{\dagger}d_{\sigma}$.

The tunneling Hamiltonian is given by
\begin{equation}
H_{T}=\sum_{\eta k\sigma}(t_{k\sigma}a_{\mathrm{\mathrm{\eta}},k\sigma}^{\dagger}d_{\sigma}+t^{*}d_{\sigma}^{\dagger}a_{\mathrm{\mathrm{\eta}},k\sigma}),
\end{equation}
where $t_{k\sigma}$ is the tunneling amplitude between the leads
$\eta$ and the QD. The definition of the shot noise is given by
\begin{widetext}
\begin{align}
S_{\eta\eta'}(t,t') & =\hbar\left\langle \{\delta\hat{I}_{\eta}(t),\delta\hat{I}_{\eta'}(t')\}\right\rangle \nonumber \\
 & =\hbar\Biggl\langle\{\hat{I}_{\eta}(t),\hat{I}_{\eta'}(t')\}\Biggr\rangle-2\hbar\Biggl\langle I_{\eta}(t)\Biggr\rangle\Biggl\langle I_{\eta'}(t')\Biggr\rangle\nonumber \\
 & =-\frac{e^{2}}{\hbar}\sum_{kk'\sigma\sigma'}\Biggl\{ t_{k\sigma}t_{k'\sigma'}\Biggl\langle a_{\eta k\sigma}^{\dagger}(t)d_{\sigma}(t)a_{'\eta'k'\sigma}^{\dagger}(t')d_{\sigma'}(t')\Biggr\rangle\nonumber \\
 & \thinspace\thinspace\thinspace\thinspace\thinspace\thinspace\thinspace\thinspace\thinspace\thinspace\thinspace\thinspace\thinspace\thinspace-t_{k\sigma}t_{k'\sigma'}^{*}\Biggl\langle a_{\eta k\sigma}^{\dagger}(t)d_{\sigma}(t)d_{\sigma'}^{\dagger}(t')a_{\eta'k'\sigma'}(t')\Biggr\rangle\nonumber \\
 & \thinspace\thinspace\thinspace\thinspace\thinspace\thinspace\thinspace\thinspace\thinspace\thinspace\thinspace\thinspace\thinspace\thinspace-t_{k\sigma}^{*}t_{k'\sigma'}\Biggl\langle d_{\sigma}^{\dagger}(t)a_{\eta k\sigma}(t)a_{\eta'k'\sigma'}^{\dagger}(t')d_{\sigma'}(t')\Biggr\rangle\nonumber \\
 & \thinspace\thinspace\thinspace\thinspace\thinspace\thinspace\thinspace\thinspace\thinspace\thinspace\thinspace\thinspace\thinspace\thinspace+t_{k\sigma}^{*}t_{k'\sigma'}^{*}\Biggl\langle d_{\sigma}^{\dagger}(t)a_{\eta k\sigma}(t)d_{\sigma'}^{\dagger}(t')a_{\eta'k'\sigma'}(t')\Biggr\rangle\Biggr\}+\mathrm{H.C.}-2\hbar\Biggl\langle I_{\eta}(t)\Biggr\rangle\Biggl\langle I_{\eta'}(t')\Biggr\rangle.
\end{align}
Using the notation for four-type of the two-particle Green's functions

\begin{align}
G_{1}^{(2)}(\tau,\tau') & =i^{2}\Biggl\langle T_{C}a_{\eta k\sigma}^{\dagger}(\tau)d_{\sigma}(\tau)a_{\eta'k'\sigma'}^{\dagger}(\tau')d_{\sigma'}(\tau')\Biggr\rangle,\\
G_{2}^{(2)}(\tau,\tau') & =i^{2}\Biggl\langle T_{C}a_{\eta k\sigma}^{\dagger}(\tau)d_{\sigma}(\tau)d_{\sigma'}^{\dagger}(\tau')a_{\eta'k'\sigma'}(\tau')\Biggr\rangle,\\
G_{3}^{(2)}(\tau,\tau') & =i^{2}\Biggl\langle T_{C}d_{\sigma}^{\dagger}(\tau)a_{\eta k\sigma}(\tau)a_{\eta'k'\sigma'}^{\dagger}(\tau')d_{\sigma'}(\tau')\Biggr\rangle,\\
G_{4}^{(2)}(\tau,\tau') & =i^{2}\Biggl\langle T_{C}d_{\sigma}^{\dagger}(\tau)a_{\eta k\sigma}(\tau)d_{\sigma'}^{\dagger}(\tau')a_{\eta'k'\sigma'}(\tau')\Biggr\rangle,
\end{align}
the shot noise can be expressed as

\begin{align}
S_{\eta\eta'}(t,t') & =\frac{e^{2}}{\hbar}\sum_{kk',\sigma\sigma'}\Biggl\{ t_{k\sigma}t_{k'\sigma'}G_{1}^{(2)>}(t,t')-t_{k\sigma}t_{k'\sigma'}^{*}G_{2}^{(2)>}(t,t')\nonumber \\
 & \thinspace\thinspace\thinspace\thinspace\thinspace\thinspace\thinspace\thinspace\thinspace\thinspace\thinspace\thinspace-t_{k\sigma}^{*}t_{k'\sigma'}G_{3}^{(2)>}(t,t')+t_{k\sigma}^{*}t_{k'\sigma'}^{*}G_{4}^{(2)>}(t,t')\Biggr\}+\mathrm{H.C.}-2\hbar\Biggl\langle I_{\eta}(t)\Biggr\rangle\Biggl\langle I_{\eta'}(t')\Biggr\rangle,
\end{align}
where $G_{i}^{(2)>}(t,t')$ can be obtained from $G_{i}^{(2)}(\tau,\tau')$
via analtyical continuation. The expression of $G_{i}^{(2)}(\tau,\tau')$
can be reduced by using the S-matrix expansion and Wick's theorem,
and more details can be found in Ref.\,\citep{PhysRevB.78.155303}.
After the reduction of$G_{i}^{(2)}$, we can obtain,

\begin{align}
G_{1}^{(2)}(\tau,\tau') & =t_{k\sigma}^{*}t_{k'\sigma'}^{*}\int\int\mathrm{d}\tau_{1}\mathrm{d}\tau_{2}G_{\eta,k\sigma\sigma}(\tau_{1},\tau)G_{\eta',k'\sigma'\sigma'}(\tau_{2},\tau')\nonumber \\
 & \thinspace\thinspace\thinspace\times\left[G_{\mathrm{QD},\sigma\sigma}(\tau,\tau_{1})G_{\mathrm{QD},\sigma'\sigma'}(\tau',\tau_{2})-G_{\mathrm{QD},\sigma\sigma'}(\tau,\tau_{2})G_{\mathrm{QD},\sigma'\sigma}(\tau',\tau_{1})\right],\\
G_{2}^{(2)}(\tau,\tau') & =-\delta_{kk'\sigma'\sigma'\eta\eta'}G_{\eta,k\sigma'\sigma}(\tau',\tau)G_{\mathrm{QD},\sigma\sigma'}(\tau,\tau')\nonumber \\
 & \thinspace\thinspace\thinspace+t_{k\sigma}^{*}t_{k'\sigma'}\int\int\mathrm{d}\tau_{1}\mathrm{d}\tau_{2}G_{\eta,k\sigma\sigma}(\tau_{1},\tau)G_{\eta',k'\sigma'\sigma'}(\tau',\tau_{2})\nonumber \\
 & \thinspace\thinspace\thinspace\times\left[G_{\mathrm{QD},\sigma\sigma}(\tau,\tau_{1})G_{\mathrm{QD},\sigma'\sigma'}(\tau_{2},\tau')-G_{\mathrm{QD},\sigma\sigma'}(\tau,\tau')G_{\mathrm{QD},\sigma'\sigma}(\tau_{2},\tau_{1})\right],\\
G_{3}^{(2)}(\tau,\tau') & =\left[G_{2}^{(2)}(\tau,\tau')\right]^{*},\\
G_{4}^{(2)}(\tau,\tau') & =\left[G_{1}^{(2)}(\tau,\tau')\right]^{*},
\end{align}
where $G_{\eta,k\sigma\sigma}(\tau_{1},\tau)=-i\Biggl\langle T_{C}a_{\eta k\sigma}(\tau_{1})a_{\eta k\sigma}^{\dagger}(\tau)\Biggr\rangle$
and $G_{\mathrm{QD},\sigma\sigma'}(\tau,\tau_{2})=-i\Biggl\langle T_{C}d_{\sigma}(\tau)d_{\sigma'}^{\dagger}(\tau_{2})\Biggr\rangle$.
The two-particle Green's functions above can be decomposed into the
connected and disconnected terms, i.e., $G_{i}^{(2)}(\tau,\tau')=G_{i,\mathrm{disc}}^{(2)}(\tau,\tau')+G_{i,\mathrm{conn}}^{(2)}(\tau,\tau')$.
The disconnected terms are given by

\begin{align}
G_{1,\mathrm{disc}}^{(2)}(\tau,\tau') & =t_{k\sigma}^{*}t_{k'\sigma'}^{*}\int\mathrm{d}\tau_{1}G_{\mathrm{QD},\sigma\sigma}(\tau,\tau_{1})G_{\eta,k\sigma\sigma}(\tau_{1},\tau^{+})\int\mathrm{d}\tau_{2}G_{\mathrm{QD},\sigma'\sigma'}(\tau',\tau_{2})G_{\eta',k'\sigma'}(\tau_{2},\tau'^{+}),\\
G_{2,\mathrm{disc}}^{(2)}(\tau,\tau') & =t_{k\sigma}^{*}t_{k'\sigma'}\int\mathrm{d}\tau_{1}G_{\mathrm{QD},\sigma\sigma}(\tau,\tau_{1})G_{\eta,k\sigma\sigma}(\tau_{1},\tau^{+})\int\mathrm{d}\tau_{2}G_{\eta',k'\sigma'\sigma'}(\tau',\tau_{2})G_{\mathrm{QD},\sigma'\sigma'}(\tau_{2},\tau'^{+}),\\
G_{3,\mathrm{disc}}^{(2)}(\tau,\tau') & =\left[G_{2,\mathrm{disc}}^{(2)}(\tau,\tau')\right]^{*},\\
G_{4,\mathrm{disc}}^{(2)}(\tau,\tau') & =\left[G_{1,\mathrm{disc}}^{(2)}(\tau,\tau')\right].^{*}
\end{align}
The analytic continuation rules give

\begin{align}
G_{1,\mathrm{disc}}^{(2)>}(t,t') & =t_{k\sigma}^{*}t_{k'\sigma'}^{*}F_{\eta,k\sigma}(t,t)F_{\eta',k'\sigma'}(t',t'),\\
G_{2,\mathrm{disc}}^{(2)>}(t,t') & =-t_{k\sigma}^{*}t_{k'\sigma'}F_{\eta,k\sigma}(t,t)F_{\eta',k'\sigma'}^{*}(t',t'),\\
G_{3,\mathrm{disc}}^{(2)>}(t,t') & =-t_{k\sigma}t_{k'\sigma'}^{*}F_{\eta,k\sigma}^{*}(t,t)F_{\eta',k'\sigma'}(t',t'),\\
G_{4,\mathrm{disc}}^{(2)>}(t,t') & =t_{k\sigma}t_{k'\sigma'}F_{\eta,k\sigma}^{*}(t,t)F_{\eta',k'\sigma'}^{*}(t',t'),
\end{align}
where
\begin{equation}
F_{\eta,k\sigma}(t,t)=\int\mathrm{d}t_{1}G_{\mathrm{QD},\sigma\sigma}^{R}(t,t_{1})G_{\eta,k\sigma\sigma}^{<}(t_{1},t)+G_{\mathrm{QD},\sigma\sigma}^{<}(t,t_{1})G_{\eta,k\sigma\sigma}^{A}(t_{1},t).
\end{equation}
The total contribution of the disconnected terms is

\begin{align}
\Biggl\langle\{\hat{I}_{\eta}(t),\hat{I}_{\eta'}(t')\}\Biggr\rangle_{\mathrm{disc}} & =\frac{e^{2}}{\hbar^{2}}\sum_{kk'\sigma\sigma'}\Biggl\{ t_{k\sigma}t_{k'\sigma'}G_{1,\mathrm{disc}}^{(2)>}(t,t')-t_{k\sigma}t_{k'\sigma'}^{*}G_{2,\mathrm{disc}}^{(2)>}(t,t')\nonumber \\
 & \thinspace\thinspace\thinspace-t_{k\sigma}^{*}t_{k'\sigma'}G_{3,\mathrm{disc}}^{(2)>}(t,t')+t_{k\sigma}^{*}t_{k'\sigma'}^{*}G_{4,\mathrm{disc}}^{(2)>}(t,t')\Biggr\}+\mathrm{H.C.}\nonumber \\
 & =2\frac{e^{2}}{\hbar^{2}}\sum_{kk'\sigma\sigma'}|t_{k\sigma}|^{2}|t_{k'\sigma'}|^{2}\left[F_{\eta,k\sigma}(t,t)+F_{\eta,k\sigma}^{*}(t,t)\right]\left[F_{\eta',k'\sigma'}(t',t')+F_{\eta',k'\sigma'}^{*}(t',t')\right].\label{eq:disc terms}
\end{align}
Note that $\Sigma_{\eta,k\sigma\sigma}^{<}=|t_{k\sigma}|^{2}G_{\eta,k\sigma\sigma}^{<}(t_{1},t)$,
the time-average current can be written as

\[
\Biggl\langle\hat{I}_{\eta}(t)\Biggr\rangle=\frac{e}{\hbar}\sum_{k\sigma}\int\mathrm{d}t_{1}\left[G_{\mathrm{QD},\sigma\sigma}^{R}(t,t_{1})\Sigma_{\eta,k\sigma\sigma}^{<}+G_{\mathrm{QD},\sigma\sigma}^{<}(t,t_{1})\Sigma_{\eta,k\sigma\sigma}^{A}\right]+\mathrm{H.C.},
\]
and hence,

\begin{equation}
\Biggl\langle\hat{I}_{\eta}(t)\Biggr\rangle\Biggl\langle\hat{I}_{\eta'}(t')\Biggr\rangle=2\frac{e^{2}}{\hbar^{2}}\sum_{kk'\sigma\sigma'}|t_{k\sigma}|^{2}|t_{k'\sigma'}|^{2}\left[F_{\eta,k\sigma}(t,t)+F_{\eta,k\sigma}^{*}(t,t)\right]\left[F_{\eta',k'\sigma'}(t',t')+F_{\eta',k'\sigma'}^{*}(t',t')\right].
\end{equation}
Therefore, the disconnected part of the shot noise (C22) and (C23)
are canceled out:
\begin{equation}
\Biggl\langle\{\hat{I}_{\eta}(t),\hat{I}_{\eta'}(t')\}\Biggr\rangle_{\mathrm{disc}}-2\Biggl\langle\hat{I}_{\eta}(t)\Biggr\rangle\Biggl\langle\hat{I}_{\eta'}(t')\Biggr\rangle=0.
\end{equation}
As the result, the remained part in the shot noise is only expressed
by the connected part by
\begin{align}
S_{\eta\eta'}(t,t') & =\hbar\Biggl\langle\{\hat{I}_{\eta}(t),\hat{I}_{\eta'}(t')\}\Biggr\rangle_{\mathrm{conn}}\nonumber \\
 & =\frac{e^{2}}{\hbar}\sum_{kk',\sigma\sigma'}\Biggl\{ t_{k\sigma}t_{k'\sigma'}G_{1,\mathrm{\mathrm{conn}}}^{(2)>}(t,t')-t_{k\sigma}t_{k'\sigma'}^{*}G_{2,\mathrm{\mathrm{conn}}}^{(2)>}(t,t')-t_{k\sigma}^{*}t_{k'\sigma'}G_{3,\mathrm{\mathrm{conn}}}^{(2)>}(t,t')+t_{k\sigma}^{*}t_{k'\sigma'}^{*}G_{4,\mathrm{conn}}^{(2)>}(t,t')\Biggr\}+\mathrm{H.C.}\nonumber \\
 & =\frac{e^{2}}{\hbar}\sum_{k,\sigma}|t_{k\sigma}|^{2}\delta_{\eta,\eta'}\left[G_{\eta,k\sigma\sigma}(t',t)G_{\mathrm{QD},\sigma\sigma}(t,t')+G_{\eta,k\sigma\sigma}(t,t')G_{\mathrm{QD},\sigma\sigma}(t',t)\right]^{>}\nonumber \\
 & \thinspace\thinspace\thinspace-\frac{e^{2}}{\hbar}\sum_{kk',\sigma\sigma'}|t_{k\sigma}|^{2}|t_{k'\sigma'}|^{2}\Biggl\{\left[\int\mathrm{d}t_{1}G_{\mathrm{QD},\sigma'\sigma}(t',t_{1})G_{\eta,k\sigma\sigma}(t_{1},t)\int\mathrm{d}t_{2}G_{\mathrm{QD},\sigma\sigma'}(t,t_{2})G_{\eta',k'\sigma'\sigma'}(t_{2},t')\right]^{>}\nonumber \\
 & \thinspace\thinspace\thinspace-\left[G_{\mathrm{QD},\sigma'\sigma}(t,t')\int\int\mathrm{d}t_{1}\mathrm{d}t_{2}G_{\eta,k\sigma\sigma}(t_{1},t)G_{\mathrm{QD},\sigma\sigma'}(t_{2},t_{1})G_{\eta',k'\sigma'\sigma'}(t',t_{2})\right]^{>}\nonumber \\
 & \thinspace\thinspace\thinspace-\left[G_{\mathrm{QD},\sigma'\sigma}(t',t)\int\int\mathrm{d}t_{1}\mathrm{d}t_{2}G_{\eta,k\sigma\sigma}(t,t_{1})G_{\mathrm{QD},\sigma\sigma'}(t_{1},t_{2},)G_{\eta',k'\sigma'\sigma'}(t_{2},t')\right]^{>}\nonumber \\
 & \thinspace\thinspace\thinspace+\left[\int\mathrm{d}t_{1}G_{\mathrm{QD},\sigma'\sigma}(t_{1},t')G_{\eta,k\sigma\sigma}(t,t_{1})\int\mathrm{d}\tau_{2}G_{\mathrm{QD},\sigma\sigma'}(t_{2},t)G_{\eta',k'\sigma'\sigma'}(t',t_{2})\right]^{>}\Biggr\}+\mathrm{H.C.}\nonumber \\
 & =\frac{e^{2}}{\hbar}\textrm{Tr}\Biggl\{\delta_{\eta,\eta'}(\Sigma_{\eta}^{>}(t',t)\widetilde{\sigma}_{z}G_{\mathrm{QD}}^{<}(t,t')\widetilde{\sigma}_{z}+G_{\mathrm{QD}}^{>}(t',t)\widetilde{\sigma}_{z}\Sigma_{\eta}^{<}(t,t')\widetilde{\sigma}_{z})\nonumber \\
 & \thinspace\thinspace\thinspace-[\int\mathrm{d}t_{1}G_{\mathrm{QD}}(t',t_{1})\Sigma_{\eta}(t_{1},t)]^{>}\widetilde{\sigma}_{z}[\int\mathrm{d}t_{2}G_{\mathrm{QD}}(t,t_{2})\Sigma_{\eta'}(t_{2},t')]^{<}\widetilde{\sigma}_{z}\nonumber \\
 & \thinspace\thinspace\thinspace+[G_{\mathrm{QD}}(t,t')]^{>}\widetilde{\sigma}_{z}[\int\int\mathrm{d}t_{1}\mathrm{d}t_{2}\Sigma_{\eta}(t_{1},t)G_{\mathrm{QD}}(t_{2},t_{1},)\Sigma_{\eta'}(t',t_{2})]^{<}\widetilde{\sigma}_{z}\nonumber \\
 & \thinspace\thinspace\thinspace+[\int\int\mathrm{d}t_{1}\mathrm{d}t_{2}\Sigma_{\eta}(t,t_{1})G_{\mathrm{QD}}(t_{1},t_{2},)\Sigma_{\eta'}(t_{2},t')]^{>}\widetilde{\sigma}_{z}[G_{\mathrm{QD}}(t',t)]^{<}\widetilde{\sigma}_{z}\nonumber \\
 & \thinspace\thinspace\thinspace-[\int\mathrm{d}t_{1}G_{\mathrm{QD}}(t_{1},t')\Sigma_{\eta}(t,t_{1})]^{>}\widetilde{\sigma}_{z}[\int\mathrm{d}t_{2}G_{\mathrm{QD}}(t_{2},t)\Sigma_{\eta'}(t',t_{2})]^{<}\widetilde{\sigma}_{z},
\end{align}
where $G_{\mathrm{QD}}$ is the $4\times$4 matrix form of the elements
$G_{\mathrm{QD},\sigma'\sigma}$ and $\Sigma_{\eta}$ is the $4\times$4
matrix form of the elements $\Sigma_{\eta,\sigma\sigma}=\sum_{k}|t_{k\sigma}|^{2}G_{\eta,k\sigma\sigma}$.
The matrix $\widetilde{\sigma}_{z}=\mathrm{diag}(1,-1,1,-1)$ describes
the different charge of electrons and holes. Finally, we apply the
convolution property of Fourier transform $\int_{-\infty}^{\infty}\mathrm{d}(t-t')e^{i\omega'(t-t')}x(t-t')y(t'-t)=\frac{1}{2\pi}\int_{-\infty}^{\infty}\mathrm{d}\omega F[x](\omega)F[y](\omega+\omega')$
to the above shot noise and obtain
\begin{align}
S_{\eta\eta'}(\omega') & =\int_{-\infty}^{\infty}\mathrm{d}(t-t')e^{i\omega'(t-t')}S_{\eta\eta'}(t-t')\nonumber \\
 & =\frac{e^{2}}{h}\int_{-\infty}^{\infty}\mathrm{d}\omega\textrm{Tr}\{\delta_{\eta,\eta'}(\Sigma_{\eta}^{>}(\omega)\widetilde{\sigma}_{z}G_{\mathrm{QD}}^{<}(\omega+\omega')\widetilde{\sigma}_{z}+G_{\mathrm{QD}}^{>}(\omega)\widetilde{\sigma}_{z}\Sigma_{\eta}^{<}(\omega+\omega')\widetilde{\sigma}_{z})\nonumber \\
 & \thinspace\thinspace-[G_{\mathrm{QD}}(\omega)\Sigma_{\eta'}(\omega)]^{>}\widetilde{\sigma}_{z}[G_{\mathrm{QD}}(\omega+\omega')\Sigma_{\eta}(\omega+\omega')]^{<}\widetilde{\sigma}_{z}-[\Sigma_{\eta}(\omega)G_{\mathrm{QD}}(\omega)]^{>}\widetilde{\sigma}_{z}[\Sigma_{\eta'}(\omega+\omega')G_{\mathrm{QD}}(\omega+\omega')]^{<}\widetilde{\sigma}_{z}\nonumber \\
 & \thinspace\thinspace+G_{\mathrm{QD}}^{>}(\omega)\widetilde{\sigma}_{z}[\Sigma_{\eta'}(\omega+\omega')G_{\mathrm{QD}}(\omega+\omega')\Sigma_{\eta}(\omega+\omega')]^{<}\widetilde{\sigma}_{z}+[\Sigma_{\eta}(\omega)G_{\mathrm{QD}}(\omega)\Sigma_{\eta'}(\omega)]^{>}\widetilde{\sigma}_{z}G_{\mathrm{QD}}^{<}(\omega+\omega')\widetilde{\sigma}_{z}\}.
\end{align}
\end{widetext}

\begin{acknowledgments}
We would like to thank Ze-Min Huang, Zhongbo Yan and Seishi Enomoto for helpful discussions.
This work is supported in part by the National Natural Science Foundation
of China under Grants No.\,11875327, the Fundamental Research Funds
for the Central Universities, and the Sun Yat-Sen University Science
Foundation.
\end{acknowledgments}

\bibliographystyle{apsrev4-1}

\end{document}